\pdfoutput=1
\documentclass[11pt]{article}

\usepackage[final]{acl} 
\usepackage{times}
\usepackage{latexsym}
\usepackage[T1]{fontenc}
\usepackage[utf8]{inputenc}
\usepackage{microtype}
\usepackage[normalem]{ulem}
\useunder{\uline}{\ul}{}

\usepackage{xspace}
\usepackage{hyperref}
\usepackage{graphicx}
\usepackage{xcolor}
\usepackage{multirow}
\usepackage{amsfonts}
\usepackage{amsmath}
\usepackage{subfigure}
\usepackage{comment}
\usepackage{booktabs} 

\usepackage{amssymb}
\usepackage{pifont}

\newcommand{\blue}[1]{\textcolor{blue}{#1}}

\newcommand{\red}[1]{\textcolor{red}{#1}}

\title{ListT5: Listwise Reranking with Fusion-in-Decoder \\ Improves Zero-shot Retrieval}

\vspace{-0.5cm}
\author{
\parbox{0.8\linewidth}{
\centering
Soyoung Yoon$^1$\thanks{~~Work done during an internship at LG AI Research.}\hspace{0.4em}\hspace{1em}
  Eunbi Choi$^2$\hspace{1em}
  Jiyeon Kim$^3$\hspace{1em}
  Hyeongu Yun$^2$\hspace{1em}
  Yireun Kim$^2$\hspace{1em}
  Seung-won Hwang$^1$\thanks{~~Corresponding author.}
  }\vspace{0.12cm}\\
  $^1$Seoul National University
  $^2$LG AI Research
  $^3$KAIST AI
\\
\texttt{\{soyoung.yoon, seungwonh\}@snu.ac.kr}
}
\vspace{-0.5cm}

\begin{document}
\maketitle

\begin{abstract}
We propose \textsc{ListT5}, a novel reranking approach based on Fusion-in-Decoder (FiD) that handles multiple candidate passages at both train and inference time. We also introduce an efficient inference framework for listwise ranking based on $m$-ary tournament sort with output caching. We evaluate and compare our model on the BEIR benchmark for zero-shot retrieval task, demonstrating that \textsc{ListT5} (1) outperforms the state-of-the-art RankT5 baseline with a notable +1.3 gain in the average NDCG@10 score, (2) has an efficiency comparable to pointwise ranking models and surpasses the efficiency of previous listwise ranking models, and (3) overcomes the lost-in-the-middle problem of previous listwise rerankers. Our code, model checkpoints, and the evaluation framework are fully open-sourced at \url{https://github.com/soyoung97/ListT5}.
\end{abstract}

\section{Introduction}
\label{sec:introduction}

Recent advancements on neural information retrievers have made significant progress in their semantic search capabilities. However, they still struggle in zero-shot or out-domain tasks where statistical retrievers such as BM25~\cite{bm25} often outperform, a crucial challenge since the setting is closely related to real-world scenarios.

Until now, the field of zero-shot reranking has been generally driven by cross-encoder models~\cite{indefense} such as MonoT5 \cite{monot5} or RankT5~\cite{rankt5}.
These models rely on \emph{pointwise} reranking of each passage, and thus lacks the ability to compare between passages relatively at inference time. This could lead to a suboptimal solution in the task of reranking, where discrimination and ordering between passages are crucial.

\begin{figure}[!ht]
{
\centering
    \includegraphics[width=0.95\columnwidth]{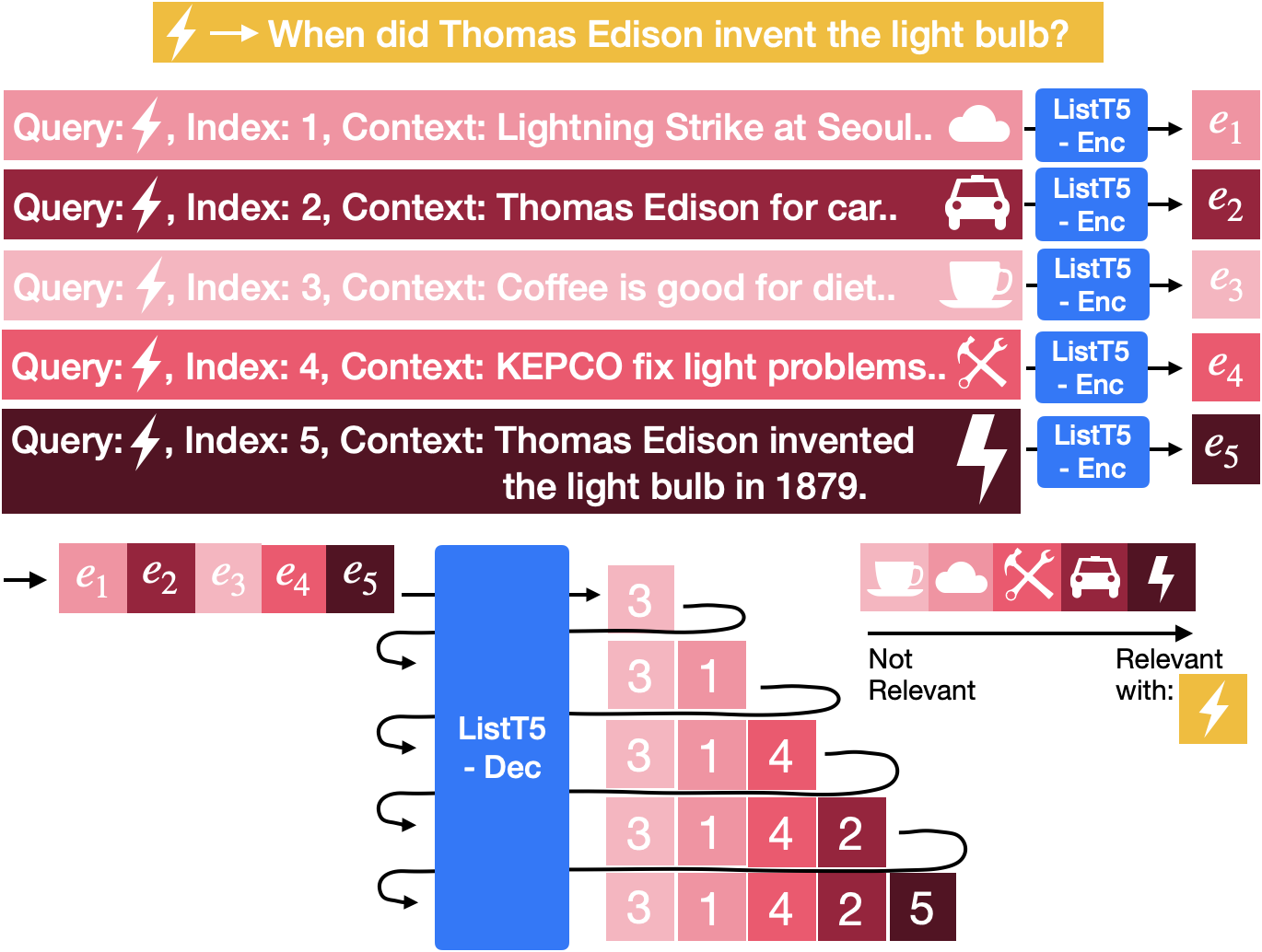}
    \caption{Operating unit of \textsc{ListT5}. \textsc{ListT5} jointly considers multiple (5) candidate passages at once using FiD, each concatenated with the query and an identifier. The output is an ordered list of the identifiers (numbers) where the most relevant passage comes at the \textit{last}.}
    \label{fig:fig_firstpage}
    \vspace{-0.5cm}
}
\end{figure}

Recently, \emph{listwise} reranking models~\cite{listwisereranking, rankgpt, distillrankgpt}, which evaluates multiple passages together, is gaining attention for its effectiveness in zero-shot retrieval. Listwise rerankers can condition on and compare multiple passages to calibrate the relevance scores better, and can reduce the inaccuracy of zero-shot predictions arising from domain shift, as theoretically supported by~\citet{xian2023learning}, and empirically evidenced by a line of works such as~\citet{listwisereranking, distillrankgpt}.

However, existing approaches that attempt listwise reranking have limitations. DuoT5~\cite{duot5} implements pairwise reranking, a form of listwise reranking as pair, which incurs quadratic time complexity to the number of candidate passages. On the other hand, listwise reranking using Large Language Models (LLMs)~\cite{rankvicuna, rankgpt, distillrankgpt, listwisereranking} sacrifice efficiency in another aspect due to large parametric model size, and also face the "Lost in the middle" problem~\cite{lostinthemiddle, foundinthemiddle}.

In this work, we present \textsc{ListT5} (Fig.~\ref{fig:fig_firstpage}), a novel listwise approach for zero-shot reranking that overcomes the aforementioned limitations.
\textsc{ListT5} jointly considers the relevancy of multiple candidate passages at both training and inference time based on Fusion-in-Decoder (FiD)~\cite{fid} architecture.
It outputs a sorted list of input passages in the \textit{increasing} order of relevance, and employs a novel \emph{tournament tree} structure at inference time for efficient sorting.
With these, \textsc{ListT5} provides the following contributions:

\textbf{(1) Computational Efficiency.} 
\textsc{ListT5} improves efficiency over previous methods in two aspects. First, for reranking top-$k$ passages given $n$ candidates, \textsc{ListT5} achieves $O(n+k\log n)$ asymptotic cost, lower than $O(n^2)$ of pairwise methods and comparable to $O(n)$ of pointwise methods (Sec.~\ref{method:time_complexity}). We empirically demonstrate such efficiency through FLOPs analysis (Sec.~\ref{results:efficiency_flops}). Second, while previous listwise reranking methods relied on LLMs to process long listwise inputs, we show that adopting FiD removes the dependency on LLMs and allows us to perform listwise reranking with smaller and more parameter efficient architectures, such as T5-base (Tab.~\ref{table/rankgpt_duot5}).

\textbf{(2) Robustness to Positional Bias.} 
In addition to efficiency, we show that \textsc{ListT5} overcomes the “Lost in the middle” problem commonly encountered in LLM-based listwise rerankers~\cite{lostinthemiddle} which tend to be positionally biased to passages presented in the first and last parts of the listwise input (Sec.~\ref{sec:results_lostinthemiddle}) and are more robust to the change in initial ordering of the passage (Sec.~\ref{sec:results_initial_ordering}). We attribute the robustness of \textsc{ListT5} to such bias to the nature of FiD, which distinguishes each input passage with identifiers, instead of positions as in LLMs.

\textbf{(3) Zero-shot Performance.}
On top of efficiency and robustness, listwise reranking with \textsc{ListT5} demonstrates strong performance in zero-shot retrieval, compared to not only the pointwise and pairwise methods, but also listwise methods based on LLMs that are argued to be specialized for zero-shot~\cite{rankvicuna, rankzephyr}. Specifically, on the BEIR benchmark for zero-shot retrieval, our approach using T5-3B surpasses state-of-the-art methods including pointwise RankT5 (+1.4 gain in NDCG@10) (Tab.~\ref{table/main_table}) and listwise RankZephyr (+1.4 gain in NDCG@10) (Tab.~\ref{table/rankgpt_duot5}). We further perform a comprehensive ablation study and find that our key components, such as tournament sort (App.~\ref{appendix/efficiency}) and generating in the increasing order of relevance, benefit zero-shot retrieval (Sec.~\ref{sec:results_design_choice}).

\section{Related Work}
\label{sec:related_work}
\subsection{Generative Models for Reranking}
\label{sec:related_work_generative}

In the reranking scenario, rather than dual encoder models~\cite{dpr} which separately encode query and passage information, models that see query and passage information jointly at inference time~\cite{sentencebert, monot5} are shown to be effective for zero-shot retrieval~\cite{indefense}. 
Among those, formulating reranking as sequence generation, such as 
conducting listwise sorting~\cite{listwisereranking, rankgpt, rankvicuna} or generating rationales~\cite{exaranker}, has shown an advantage in application to zero-shot retrieval by leveraging the language model's auto-regressive generation capabilities. Specifically, a series of studies that use the encoder-decoder architecture of T5 (Sec.~\ref{related_work_t5}), and applying zero-shot reranking with LLMs (Sec.~\ref{sec:related_work_llm}), or viewing reranking as autoregressive text generation problem~\cite{wang2024large} has been successful. 



\subsection{Listwise Reranking for T5}
\label{related_work_t5}
MonoT5~\cite{monot5} leaverages pre-defined token probabilities (i.e., true/false) as relevance scores at inference time. RankT5~\cite{rankt5} introduces listwise training loss but both MonoT5 and RankT5 performs pointwise reranking at inference time. Built on top of MonoT5, DuoT5~\cite{duot5}, implements pairwise reranking, showing superior performance. However, DuoT5 only sees two passages at each prediction, and is inefficient 
as it has to run prediction on every pair of candidate passages (thus requiring $n^2-n$ predictions to rerank $n$ passages).

\subsection{Listwise Reranking with LLMs}
\label{sec:related_work_llm}
A line of work \cite{rankgpt,rankvicuna, listwisereranking, rankzephyr, llms-effective-rankers} implements listwise reranking to LLMs, usually in a format of taking multiple (20) passages as input with the sliding window approach. 
However, 
this format presents a challenge due to the monolithic and lengthy input size. 
The model must process all tokens at once, limiting its applicability to LLMs trained with large context lengths. 
The lengthy input size also leads to the "lost in the middle" problem~\cite{lostinthemiddle} for LLMs, exhibiting strong positional bias to the information in the first and last parts of long inputs, failing to comprehend relevant information in the middle. This problem is also prevalent in listwise reranking~\cite{foundinthemiddle}.
Furthermore, this approach comes at the cost of sacrificing computational efficiency in the already computation-intensive reranking scenarios.

\subsection{Listwise Reranking with $m$-ary Units}
\citet{ai2019learning} is in a similar spirit to ours, as they define a basic ranking unit on $m$ elements and extends it to rank the list of $n(\gg m)$ elements.
However, while they consider an intractable number of all $m!$ orderings of all $m$-sized subsets and then employ Monte Carlo sampling for approximation, we use $m$-ary tournament sort algorithm which is complete and efficient by construction (Sec.~\ref{method/extension_torunament_sort}).


\begin{figure}[!t]
{
\centering
    \includegraphics[width=\columnwidth]{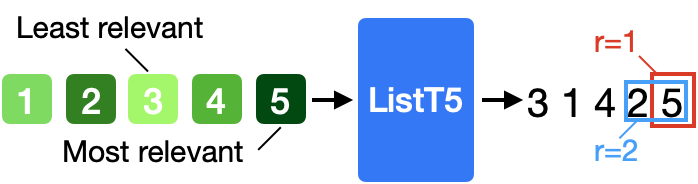}
    \caption{
    Two variants of \textsc{ListT5}, ($r$=1) and ($r$=2). The underlying model is identical and only the inference method varies. ($r$=1) keeps only the top 1 relevant index from the output, and ($r$=2) keeps top 2 relevant indices.
    \label{fig:fig_top1top2}
    }
}
\end{figure}
\section{Methods}
\label{sec:methods}
Our method \textsc{ListT5} for listwise reranking has two components: \textbf{(1)} the basic operating unit that takes a fixed number of $\boldsymbol{m}$ passages and ranks the top-$\boldsymbol{r}$ based on FiD, and \textbf{(2)} an extension of this basic unit that takes full $\boldsymbol{n}$ passages and ranks the top-$\boldsymbol{k}$ based on tournament sort. We describe the basic unit in Sec.~\ref{method/basic_operating_unit} and its extension in Sec.~\ref{method/extension_torunament_sort}.

\subsection{Basic Operating Unit ($\boldsymbol{m \to r}$)}
\label{method/basic_operating_unit}
The basic operating unit of \textsc{ListT5} processes a fixed number of $m$ input passages and outputs top-$r$ passages ($m\to r$), as in Fig.~\ref{fig:fig_top1top2}.
We develop the basic unit upon the Fusion-in-Decoder (FiD) architecture~\cite{fid}, which modifies the Encoder-Decoder structure of T5~\cite{t5} to process listwise inputs efficiently.
Specifically, given a query $\mathbf{q}$ and a list of $m$ candidate passages $[\mathbf{p}_1, ..., \mathbf{p}_m]$, our basic unit concatenates each passage $\mathbf{p}_i$ with its index identifier $i$ and the query $\mathbf{q}$ in the following format, and then feeds them to Encoder to obtain their full sequence token embedding $\mathbf{h}_{i}$:

\vspace{-0.5cm}
\begin{equation}
\mathbf{h}_{i} = \operatorname{Enc}(\text{"Question: }\mathbf{q}\text{, Index: }i\text{, Context: }\mathbf{p}_{i}\text{"})
\label{eq:listt5_input}
\end{equation}

In Eq.~\eqref{eq:listt5_input}, notice that each passage $\mathbf{p}_i$ is encoded separately.
Listwise reasoning upon multiple passages $\mathbf{p}_1, ..., \mathbf{p}_m$ is handled by the decoder.
Specifically, the encoded representations $[\mathbf{h}_1, ..., \mathbf{h}_m]$ are concatenated to a single sequence and fed to the decoder, which performs listwise ranking and auto-regressively generates an ordered sequence of the passage indices (identifiers) $[i'_{1}, ..., i'_{m}]$ in the \emph{increasing} order of relevance:
\begin{align}
[i'_{1}, ..., i'_{m}] &= \operatorname{Dec}(\operatorname{concat}[\mathbf{h}_{1}, ..., \mathbf{h}_{m}]),
\nonumber\\
\text{ where }&\operatorname{rel}(\mathbf{q}, \mathbf{p}_{i'_{1}}) < ... < \operatorname{rel}(\mathbf{q}, \mathbf{p}_{i'_{m}})\label{eq:listt5_output}
\end{align}
Importantly, in Eq.~\eqref{eq:listt5_output}, notice that the decoder generates output passages from the \emph{least} relevant one $i_1'$ to the \emph{most} relevant one $i_m'$.
This is different from previous listwise rerankers (Sec.~\ref{sec:related_work_llm}), which generate from the most relevant passage index.
Our rationale is that generating from the least relevant to the most relevant can be beneficial, as it may act a reasoning chain that progressively eliminates irrelevant passages to deduce the most relevant passages.
One can imagine solving a confusing multiple-choice question; crossing out the options that are definitely not the answer until the last remaining option serves as a good strategy.

After the decoder ranks all $m$ input passages, we decide the number $r$ of relevant passages we will keep as the output of the basic operating unit.
Namely, \textbf{\textsc{ListT5} ($\mathbf{r}$ = 1)} keeps one, so only $[\mathbf{p}_{i'_{m}}]$ is selected as the output. \textbf{\textsc{ListT5} ($\mathbf{r}$ = 2)} keeps two, selecting $[\mathbf{p}_{i'_{m}},\mathbf{p}_{i'_{m-1}}]$ as the output.
Larger choices of $r$ are possible in principle, but since we have observed saturation of performance for $r>2$ in early trials, we only use the $r$ = 1 and $r$ = 2 variants of the basic unit in our experiments.
For $m$, we set its value to $5$ in our experiments. Ablations on the choice of $m$ can be found at App.~\ref{appendix/ablation_topk}.


\subsection{Extension with Tournament Sort ($\boldsymbol{n \to k}$)}
\label{method/extension_torunament_sort}

While the basic operating unit of \textsc{ListT5}~(Sec.~\ref{method/basic_operating_unit}) reranks a fixed number of $m\,(=5)$ passages, our end goal is to rerank $k$ passages given a much larger number of $n\,(\gg m)$ candidate passages.
This necessitates an algorithm to extend the basic unit ($m\to r$) to full reranking ($n\to k$).
For the similar purpose, previous listwise rerankers (Sec.~\ref{sec:related_work_llm}) has mostly used the sliding window algorithm.
In these, an operating unit $m\to r$, typically an LLM, is slided over the $n$ candidate passages, and at each step $m$ passages are locally reranked and reordered. After the sliding is done, the top $k$ passages are used as the output of global reranking.

However, we observe that sliding window has a drawback.
Since a window of size $m$ can only "cache" up to $m$ passages, full reranking $n\to k$ is bound to be inaccurate when $k$ is set to be $>m$.
In such cases, it is necessary to run the whole sliding over the entire $n$ passages multiple times (\emph{e.g.}, $\lceil\frac{k}{m}\rceil$ times).
Each sliding run would cost $\mathcal{O}(n)$ asymptotically, which can be computationally demanding given $n\gg m$.
Therefore, for \textsc{ListT5}, we opt into developing a novel algorithm that does not require multiple evaluations over the entire $n$ passages.

\begin{figure*}[!t]
{
\centering
    \includegraphics[width=\textwidth]{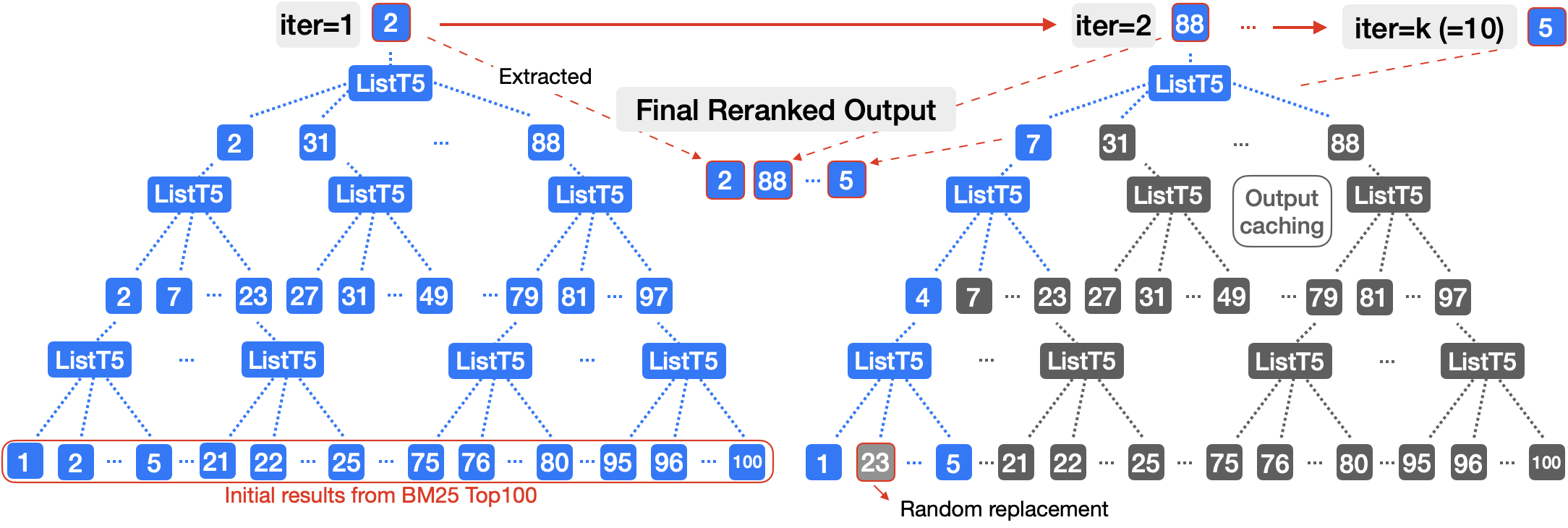}
    \caption{Illustration of our inference framework using $m$-ary tournament sort, with \textsc{ListT5} ($r$ = 1) as the basic unit. Given $n$ candidates, we can order top-$k$ most relevant passages in $O(n+k\log{n})$ asymptotic complexity. We can use either ($r$ = 1) or ($r$ = 2) for the basic unit, but the uppermost unit always outputs 1 ($r$ = 1). We fix $m$ to 5 in our experiments. Full illustration at Appendix Fig.\ref{fig:fig_appendix_inference}.
    }
    \label{fig:inference_tree}
}
\end{figure*}




Our inference algorithm for \textsc{ListT5} that extends the basic unit ($m\to r$) to full reranking ($n\to k$) is inspired by the \emph{tournament sort} algorithm~\cite{tournamentsort} (App.~\ref{appendix/tournament_tree}).
Given $n$ inputs, tournament sort ranks top-$k$ by \textbf{(1)} constructing a binary tournament tree of $n$ inputs where the root is the top-1 element, and \textbf{(2)} running $k$ iterations of extracting the root and re-doing the tournament with the remaining inputs.
Here, a crucial property is that once the tournament tree has been constructed, re-doing the tournament only requires recomputing a \emph{single} path that traverses from a leaf to the root. As a result, most of the nodes and edges can be cached and reused over the $k$ iterations, and each iteration only costs $\mathcal{O}(\log_2 n)$ asymptotically.
Unlike sliding window, multiple evaluations over the entire $n$ inputs is avoided.


Based on tournament sort, our inference algorithm $n\to k$ for \textsc{ListT5} is defined by 
\textbf{(1)} extending from binary to $m$-ary tournament trees, and \textbf{(2)} invoking the basic unit $m\to r$ (Sec.~\ref{method/basic_operating_unit}) at each node of the tournament tree.
Specifically, given an arbitrary number of $n$ passages, we perform a bottom-up comparison using the basic unit to construct an $m$-ary tournament tree.
Then, we run $k$ iterations of extracting the top-1 passage at the root and re-doing the tournament by recomputing a single path from the replaced leaf to root.
In this way, the model can rank top-$k$ passages from an arbitrary number of $n$ passages while inheriting the efficiency of tournament sort.
Fig.~\ref{fig:inference_tree} illustrates our inference algorithm using \textsc{ListT5} ($r$ = 1) as the basic unit.
Further details can be found at App.~\ref{appendix/inference}.

\paragraph{Asymptotic Complexity.}
\label{method:time_complexity}
\begin{table}
\resizebox{\columnwidth}{!}
{
\begin{tabular}{@{}l|l|l@{}}
\toprule
Method Name                                  
& \begin{tabular}[c]{@{}l@{}}Reranking\\ Method \end{tabular} & Complexity                             \\ \midrule
\begin{tabular}[c]{@{}l@{}}MonoT5~\cite{monot5},\\ RankT5~\cite{rankt5}\end{tabular} & Pointwise                                                  & $\mathcal{O}(n)$                       \\ \midrule
DuoT5~\cite{duot5}                                                    & Pairwise                                                   & $\mathcal{O}(n^2)$                     \\ \midrule
ListT5 (Ours)                                                  & Listwise                                                   & \textbf{$\mathcal{O}(n + k\log{n})$} \\ \bottomrule
\end{tabular}
}

\caption{Comparison of the computational complexity of different ranking methods on obtaining top-$k$ passages from $n$ candidate passages.
The base of the log function is $m$. \textsc{ListT5} achieves $\mathcal{O}(n + k\log{n})$, or $\mathcal{O}(n + n\log{n})$ when $n$ = $k$, which is way better than pairwise models, competitive with pointwise models, and is in fact the best possible (worst-case asymptotic) complexity for sorting algorithms based on comparisons~\cite{ren2018pac}\footnotemark.}
\label{table/inference_time_complexity}
\end{table}

We now present an asymptotic analysis of the cost of listwise reranking $n\to k$ with \textsc{ListT5} using a basic unit $m\to r$.
At the first iteration of tournament sort, we need to construct the full tournament tree, which amounts to $n(\frac{1}{m}+\frac{r}{m^2}+\frac{r^2}{m^3}+...)=\frac{n}{m}\frac{m}{m-r}$ evaluations which is $\mathcal{O}(n)$.
In each iteration afterwards, we only need to recompute a single path from the leaf to root to compute the top-1, resulting in an asymptotic cost of $\mathcal{O}(\log_mn)$.
With $k$ iterations, \textsc{ListT5} achieves $\mathcal{O}(n + k\log_{m} n)$ asymptotic cost for reranking top-$k$ from $n$ candidate passages.
If we compare this complexity with others (Tab.~\ref{table/inference_time_complexity}\footnotetext{\url{https://en.wikipedia.org/wiki/Comparison_sort}}), we can see that \textsc{ListT5} is more efficient than pairwise reranking models of $\mathcal{O}(n^2)$~\cite{duot5}, and comparable to pointwise models of $\mathcal{O}(n)$~\cite{monot5, rankt5}. 

\begin{table*}[t!]

\centering
\resizebox{\textwidth}{!}{

\begin{tabular}{@{}l|ccccccc|ccc@{}}
\toprule
 & \multicolumn{7}{c|}{BM25 Top-100} & \multicolumn{3}{c}{BM25 Top-1000} \\  \cmidrule(lr){2-8} \cmidrule(lr){9-11}
 & Initial & \begin{tabular}[c]{@{}c@{}}MonoT5\\ -base\end{tabular} & \begin{tabular}[c]{@{}c@{}}RankT5\\ -base\end{tabular} & \multicolumn{1}{c|}{\begin{tabular}[c]{@{}c@{}}ListT5\\ -base\\ (r=2)\end{tabular}} & \begin{tabular}[c]{@{}c@{}}MonoT5\\ -3B\end{tabular} & \begin{tabular}[c]{@{}c@{}}RankT5\\ -3B\end{tabular} & \begin{tabular}[c]{@{}c@{}}ListT5\\ -3B\\ (r=2)\end{tabular} & \begin{tabular}[c]{@{}c@{}}MonoT5\\ -base\end{tabular} & \begin{tabular}[c]{@{}c@{}}RankT5\\ -base\end{tabular} & \begin{tabular}[c]{@{}c@{}}ListT5\\ -base\\ (r=2)\end{tabular} \\ \midrule
TREC-COVID & 59.5 & \textbf{78.3} & 77.7 & \multicolumn{1}{c|}{\textbf{78.3}} & 79.8 & 81.7 & \textbf{84.7} & 78.3 & 79.1 & \textbf{82.1} \\
NFCorpus & 32.2 & \textbf{35.7} & 35.1 & \multicolumn{1}{c|}{35.6} & 37.3 & \textbf{37.4} & 37.7 & \textbf{36.1} & 35.3 & \textbf{36.1} \\
BioASQ & 52.2 & 55.3 & \textbf{58.2} & \multicolumn{1}{c|}{56.4} & 57.5 & \textbf{58.3} & \textbf{58.3} & 52.6 & \textbf{57.6} & 55.0 \\
NQ & 30.5 & 52.1 & \textbf{53.2} & \multicolumn{1}{c|}{53.1} & 56.4 & \textbf{57.8} & 56.2 & 55.9 & \textbf{57.6} & 57.5 \\
HotpotQA & 63.3 & 71.2 & \textbf{72.8} & \multicolumn{1}{c|}{72.6} & 74.3 & 74.8 & \textbf{75.6} & 70.9 & \textbf{73.8} & 73.6 \\
FiQA-2018 & 23.6 & 39.2 & 39.2 & \multicolumn{1}{c|}{\textbf{39.6}} & \textbf{46.0} & 45.2 & 45.1 & 41.2 & 41.1 & \textbf{41.8} \\
Signal-1M (RT) & 33.0 & 32.0 & 30.8 & \multicolumn{1}{c|}{\textbf{\red{33.5}}} & 32.2 & 31.9 & \textbf{33.8} & 29.3 & 28.6 & \textbf{30.9} \\
TREC-NEWS & 39.5 & 48.0 & 45.4 & \multicolumn{1}{c|}{48.5} & 48.3 & 49.5 & \textbf{53.2} & 47.8 & 45.9 & \textbf{50.9} \\
Robust04 & 40.7 & 53.4 & \textbf{54.3} & \multicolumn{1}{c|}{52.1} & \textbf{58.5} & 58.3 & 57.8 & 55.4 & \textbf{57.2} & 54.7 \\
Arguana & 40.8 & 34.4 & 35.5 & \multicolumn{1}{c|}{\textbf{\red{48.9}}} & 46.8 & 37.4 & \textbf{50.6} & 24.2 & 26.6 & \textbf{46.9} \\
Touche-2020 & 44.2 & 29.6 & \textbf{37.1} & \multicolumn{1}{c|}{33.4} & 32.5 & \textbf{38.8} & 33.6 & 26.4 & \textbf{37.0} & 31.5 \\
CQADupStack & 30.0 & 38.6 & 37.0 & \multicolumn{1}{c|}{\textbf{38.8}} & 41.3 & 40.3 & \textbf{42.1} & 40.1 & 38.1 & \textbf{40.5} \\
Quora & 78.9 & 84.6 & 83.3 & \multicolumn{1}{c|}{\textbf{\red{86.4}}} & 84.0 & 83.6 & \textbf{86.9} & 84.2 & 82.9 & \textbf{86.4} \\
DBPedia & 31.8 & 42.8 & 43.7 & \multicolumn{1}{c|}{43.7} & 44.8 & \textbf{45.0} & 46.2 & 43.1 & \textbf{45.1} & 44.9 \\
SCIDOCS & 14.9 & 16.7 & 16.8 & \multicolumn{1}{c|}{\textbf{17.6}} & 19.0 & 18.9 & \textbf{19.5} & 17.0 & 17.1 & \textbf{18.0} \\
FEVER & 65.2 & 78.4 & 77.6 & \multicolumn{1}{c|}{\textbf{79.8}} & 80.0 & 79.8 & \textbf{82.0} & 77.9 & 77.8 & \textbf{81.0} \\
Climate-FEVER & 16.5 & 23.1 & 21.2 & \multicolumn{1}{c|}{\textbf{24.0}} & \textbf{26.2} & 24.5 & 24.8 & 23.3 & 20.6 & \textbf{24.9} \\
SciFact & 67.9 & 73.1 & 73.5 & \multicolumn{1}{c|}{\textbf{74.1}} & 76.3 & \textbf{77.1} & 77.0 & 73.3 & 73.6 & \textbf{74.9} \\ \midrule
Average & 42.5 & 49.3 & 49.6 & \multicolumn{1}{c|}{\textbf{50.9}} & 52.3 & 52.2 & \textbf{53.6} & 48.7 & 49.7 & \textbf{51.8} \\ \bottomrule
\end{tabular}

}
\caption{Comparison of \textsc{ListT5} against \textbf{pointwise} ranking models, on BEIR, in NDCG@10.
MS MARCO-Top1000 results for 3B models are omitted due to their long inference time. \textsc{ListT5}-base excels in both performance and model size (than MonoT5-3B or RankT5-3B) for datasets in \red{red}.}
\label{table/main_table}

\end{table*}

\begin{table*}
    \resizebox{\textwidth}{!}
    {
\renewcommand{\arraystretch}{1.05}
\begin{tabular}{@{}l|cc|cccccccc|c|c@{}}
\toprule
 &
  \begin{tabular}[c]{@{}c@{}}TREC-\\ DL19\end{tabular} &
  \begin{tabular}[c]{@{}c@{}}TREC-\\ DL20\end{tabular} &
  \begin{tabular}[c]{@{}c@{}}TREC-\\ COVID\end{tabular} &
  \begin{tabular}[c]{@{}c@{}}NFC-\\ orpus\end{tabular} &
  \begin{tabular}[c]{@{}c@{}}Signal-\\ 1M (RT)\end{tabular} &
  \begin{tabular}[c]{@{}c@{}}TREC-\\ NEWS\end{tabular} &
  \begin{tabular}[c]{@{}c@{}}Robu-\\ st 04\end{tabular} &
  \begin{tabular}[c]{@{}c@{}}Touche-\\ 2020\end{tabular} &
  \begin{tabular}[c]{@{}c@{}}DBP-\\ edia\end{tabular} &
  \begin{tabular}[c]{@{}c@{}}Sci-\\ Fact\end{tabular} &
  \begin{tabular}[c]{@{}c@{}}Avg (In-\\ domain)\end{tabular} &
  \begin{tabular}[c]{@{}c@{}}Avg\\ (BeIR)\end{tabular} \\ \midrule
DuoT5-base &
  71.4 &
  67.4 &
  \textbf{80.1} &
  35.0 &
  31.4 &
  \textbf{49.1} &
  49.6 &
  31.8 &
  \textbf{43.9} &
  69.6 &
  69.4 &
  48.8 \\
  \textsc{ListT5}-base ($r = 2$) &
  \textbf{71.8} &
  \textbf{68.1} &
  78.3 &
  \textbf{35.6} &
  \textbf{33.5} &
  48.5 &
  \textbf{52.1} &
  \textbf{33.4} &
  43.7 &
  \textbf{74.1} &
  \textbf{70.0} &
  \textbf{49.9} \\ \midrule
RankGPT (GPT3.5) &
  65.8 &
  62.9 &
  76.7 &
  35.6 &
  32.1 &
  48.9 &
  50.6 &
  \textbf{36.2} &
  44.5 &
  70.4 &
  64.4 &
  49.4 \\
RankVicuna-7b &
  68.9 &
  66.1 &
  80.5 &
  33.2 &
  \textbf{34.2} &
  46.9 &
  48.9 &
  33.0 &
  44.4 &
  70.8 &
  67.5 &
  49.0 \\
RankZephyr-7b &
  \textbf{73.9} &
  \textbf{70.9} &
  84.0 &
  36.7 &
  31.8 &
  52.6 &
  54.3 &
  33.8 &
  44.6 &
  74.9 &
  \textbf{72.4} &
  51.6 \\
  \textsc{ListT5}-3B ($r = 2$) &
  71.8 &
  69.1 &
  \textbf{84.7} &
  \textbf{37.7} &
  33.8 &
  \textbf{53.2} &
  \textbf{57.8} &
  33.6 &
  \textbf{46.2} &
  \textbf{77.0} &
  70.5 &
  \textbf{53.0} \\ \bottomrule
\end{tabular}

}
\caption{Comparison of \textsc{ListT5} against \textbf{listwise} (RankGPT, RankVicuna, RankZephyr) and \textbf{pairwise} (DuoT5) ranking models, in NDCG@10. Initial results are from BM25 Top100. The scores of RankGPT (with GPT-3.5-turbo-0301) are from the RankGPT paper, and the best scores are in bold. DuoT5 is applied on top 50 passages reranked by MonoT5. \textsc{ListT5} excels the pairwise counterparts as well as the previous listwise ranking models on the selected subset of BEIR.}
\label{table/rankgpt_duot5}
\end{table*}

\section{Experiments}
\label{sec:experiments}
\subsection{Overview}
\label{sec:results-and-discussion}
In this section, after explaining the training and evaluation details, we present three main results in the following order: (1) General performance and efficiency comparison of our models with respect to pointwise and listwise ranking models at Sec.~\ref{sec:results_performance}, (2) Analyzing the lost-in-the-middle problem by measuring the robustness to positional bias between \textsc{ListT5} and listwise ranking models at Sec.~\ref{sec:results_lostinthemiddle}, and (3) Ablation experiments on the design choice at Sec.~\ref{sec:results_design_choice}. All results are based on the T5-base model except explicitly mentioning 3B.


\subsection{Training}
To train \textsc{ListT5}, we use the official \textit{train} subset of the MS MARCO \cite{msmarco} passage ranking dataset.\footnote{\url{https://microsoft.github.io/msmarco/Datasets}} The dataset consists of 532,761 distinct queries and 8.8M passages, with binary annotations of relevancy. Since the ordering of relevancy between negative passages is not provided, we use a dual-encoder retrieval model, specifically COCO-DR large \cite{cocodr}, to retrieve Top-1000 passages and label relevance scores of negatives for the training dataset of MS MARCO. Experiments using GTR \cite{gtr}, the same model used to build the training data for \citet{rankt5}, are reported at Appendix.~\ref{appendix/gtr}. 
For each query-positive passage pair, we randomly sample 20 groups of $m - 1$ distinct negative passages. We randomly shuffle the positive and negative passages and assign identifiers $\{1, ..., m\}$ to them to form each training data.


We train T5 with two different model sizes: base and 3B. We conduct grid search (App.~\ref{appendix/hyperparameter}) on the subset of BEIR on the learning rate and number of steps in a reasonable range and report the best scores. As a result, we report the T5-base model trained for 20k steps with a learning rate of $1 \times 10^{-4}$ and T5-3B for 3k steps with a learning rate of $1 \times 10^{-5}$, both with linear learning rate scheduler with 10\% warmup and an effective batch size of 256 in bfloat16 precision. Maximum input length is set to 230 and 128, and the maximum output length to 7 for T5-base and T5-3B, respectively. With DeepSpeed stage 2 optimization, training T5-base took approximately 8 hours on 4$\times$NVIDIA RTX A6000 GPUs with 48GB, while T5-3B training took 3 hours on 8$\times$NVIDIA A100 GPUs with 40GB.

\subsection{Evaluation}
We measure the evaluation performance on the official dev set of MS MARCO (passage ranking dataset), TREC-DL19, and TREC-DL20 \cite{trec20} for in-domain performance, and the BEIR benchmark \cite{beir} for zero-shot out-domain performance. BEIR contains 18 diverse sets of domains and tasks. We download the dumped index of the Top-100 and Top-1000 retrieved passages by BM25 from the official Pyserini repository. Reranking results using advanced first-stage retrieval models (e.g., COCO-DR) are reported at App.~\ref{appendix/result_cocodr}. 
We use the evaluation metric of the Normalized Discounted Cumulative Gain@10 (NDCG@10)~\cite{ndcg}, the official evaluation metric for BEIR~\cite{beir}. All reported scores are rounded to the nearest tenth. We also report the Mean Reciprocal Rank@10 (MRR@10) scores at App. Tab.~\ref{table/appendix_mrr}. 
For reproducibility, we include the evaluation details including details to run baseline models and links for dataset download at App.~\ref{appendix/evaluation}.




\begin{figure}
{
\centering
    \includegraphics[width=\columnwidth]{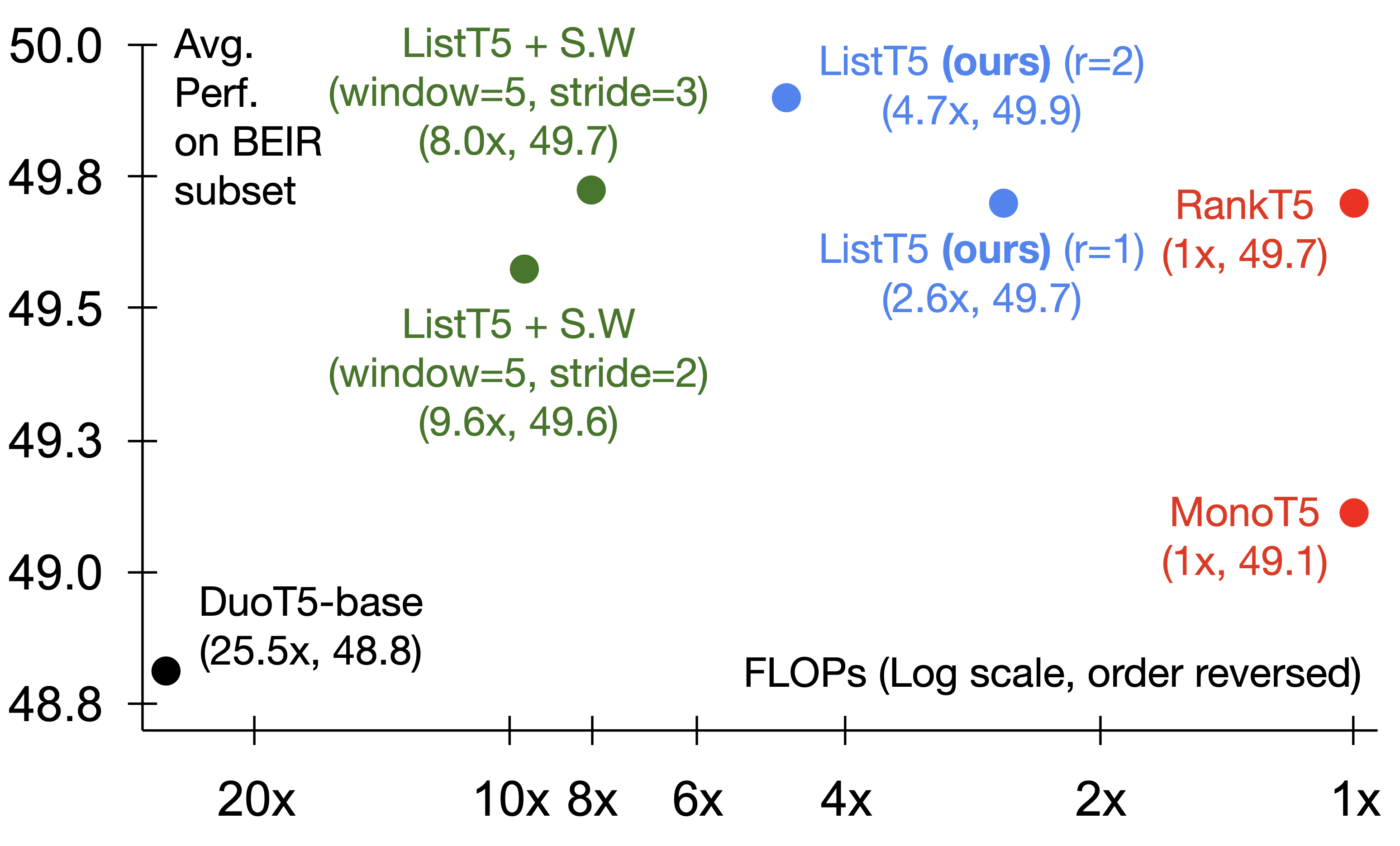}
    \caption{Real-time FLOPs comparison of the models on T5-base, including DuoT5 and the sliding window variants of \textsc{ListT5}. The reported BEIR performance is averaged from a subset of BEIR, same as in Tab.~\ref{table/rankgpt_duot5}.}
    \label{fig:fig_placeholder_flops}
}
\end{figure}

\begin{table*}

\resizebox{\textwidth}{!}
{

\begin{tabular}{@{}l|ccccccc|ccccccc@{}}
\toprule
 &
  \multicolumn{7}{c|}{TREC-COVID} &
  \multicolumn{7}{c}{FiQA} \\ \midrule
 &
  \multicolumn{6}{c|}{\begin{tabular}[c]{@{}c@{}}Accuracy when positive passage\\ is at index \# :\end{tabular}} &
  \multirow{2}{*}{\begin{tabular}[c]{@{}c@{}}Aggrement \\ ratio ($\uparrow$)\end{tabular}} &
  \multicolumn{6}{c|}{\begin{tabular}[c]{@{}c@{}}Accuracy when positive passage\\ is at index \#:\end{tabular}} &
  \multirow{2}{*}{\begin{tabular}[c]{@{}c@{}}Aggrement \\ ratio ($\uparrow$)\end{tabular}} \\ \cmidrule(r){1-7} \cmidrule(lr){9-14}
 &
  1 &
  2 &
  3 &
  4 &
  5 &
  \multicolumn{1}{c|}{Std. ($\downarrow$)} &
   &
  1 &
  2 &
  3 &
  4 &
  5 &
  \multicolumn{1}{c|}{Std. ($\downarrow$)} &
   \\ \midrule
GPT-3.5 &
  81.6 &
  63.3 &
  75.5 &
  67.3 &
  61.2 &
  \multicolumn{1}{c|}{7.7} &
  55.1 &
  88.3 &
  68.1 &
  78.7 &
  65.9 &
  75.8 &
  \multicolumn{1}{c|}{8.0} &
  62.1 \\
GPT-4 &
  95.9 &
  83.7 &
  73.5 &
  77.6 &
  71.4 &
  \multicolumn{1}{c|}{8.8} &
  69.4 &
  94.6 &
  90.5 &
  84.4 &
  86.8 &
  84.8 &
  \multicolumn{1}{c|}{3.9} &
  82.8 \\
DuoT5 &
  91.3 &
  76.0 &
  - &
  - &
  - &
  \multicolumn{1}{c|}{7.6} &
  79.6 &
  89.9 &
  76.9 &
  - &
  - &
  - &
  \multicolumn{1}{c|}{6.5} &
  78.1 \\
\textsc{ListT5} &
  93.9 &
  87.8 &
  83.7 &
  85.7 &
  81.6 &
  \multicolumn{1}{c|}{\textbf{4.2}} &
  \textbf{83.7} &
  85.3 &
  85.6 &
  82.2 &
  83.3 &
  82.6 &
  \multicolumn{1}{c|}{\textbf{1.4}} &
  \textbf{90.4} \\ \bottomrule
\end{tabular}

}

\caption{Robustness to the position of the positive passage in the input, on TREC-COVID and FiQA. GPT-3.5, GPT-4, DuoT5, and \textsc{ListT5} stands for GPT-3.5-turbo-1106, GPT-4-0613, DuoT5-base, and \textsc{ListT5}-base, respectively. Using the FiD structure effectively mitigates the problem of the positional bias of positive passages, showing lowest standard deviation and highest agreement ratio.}
\label{table/permutation_consistency_twocolumn}

\end{table*}

\subsection{Zero-shot performance and efficiency.}
\label{sec:results_performance}
\textbf{Performance.} We measure and compare the performance of \textsc{ListT5} against \textit{pointwise} ranking models (MonoT5~\cite{monot5}, RankT5~\cite{rankt5}) in Tab.~\ref{table/main_table}, and against \textit{pairwise and listwise} ranking models (DuoT5~\cite{duot5}, RankVicuna~\cite{rankvicuna}, RankZephyr~\cite{rankzephyr}) in Tab.~\ref{table/rankgpt_duot5}. \textsc{ListT5} ($r$ = 2) achieves an average of +1.3 point gain on NDCG@10 than RankT5 for reranking on BM25 Top-100. Also, the gain from listwise reranking improves even further when we rerank from BM25 Top-100 to Top-1000. While pointwise models show small performance difference from BM25 Top-100, \textsc{ListT5} improves from 50.9 to 51.8, additional 0.9 \% gain from BM25 Top-100. 
Additionally, \textsc{ListT5} also excels on 3B model variants, outperforming RankVicuna-7b and RankZephyr-7b on the selected subset of BEIR. We additionally provide qualitative analysis at App.~\ref{appendix/qualitative}.

\noindent \textbf{Efficiency in FLOPs.}
\label{results:efficiency_flops}
In addition to comparing the computational complexity in Sec.~\ref{method:time_complexity}, we also conduct real-time measurements of floating point operations (FLOPs) using the FlopsProfiler of the DeepSpeed library\footnote{\url{https://deepspeed.readthedocs.io/en/latest/flops-profiler.html}}. We measure the FLOPs needed to rerank Top-10 passages from BM25-Top100 candidate passages for TREC-DL19 (43 queries, maximum input length of 256) on T5-base, run the profiler, and report the relative FLOPs when we set the FLOPs of MonoT5 (pointwise model) as 1\footnote{The exact value of FLOPs for MonoT5 in this setup was 229,732,539,806,400.}. We also report the performance and FLOPs of the sliding window variants of \textsc{ListT5}, with details at App.~\ref{appendix/efficiency}. The results in Fig.~\ref{fig:fig_placeholder_flops} shows that \textsc{ListT5} ($r$=1) and ($r$=2) are comparable with MonoT5 and RankT5 (pointwise ranking models), and much more efficient than DuoT5 (pairwise ranking models). Also, tournament sort uses output caching to lower down FLOPs than the sliding window variants, with better performance, on the real-time FLOPs calculation (App.~\ref{appendix/efficiency}). Additionally, we can \textit{parallize} computation of the same level (e.g., leaf node) within query for tournament sort, while the sliding window approach should be computed sequentially.  Even better, the \textsc{ListT5} model is \textit{flexible} - according to their needs, users can have the choice to select between the trade-off of inference speed ($r$=1) and performance ($r$=2), an option not available for other models.

\subsection{Robustness to Positional Bias}
\label{sec:results_lostinthemiddle}

\begin{figure}[!t]
{
\centering
    \includegraphics[width=\columnwidth]{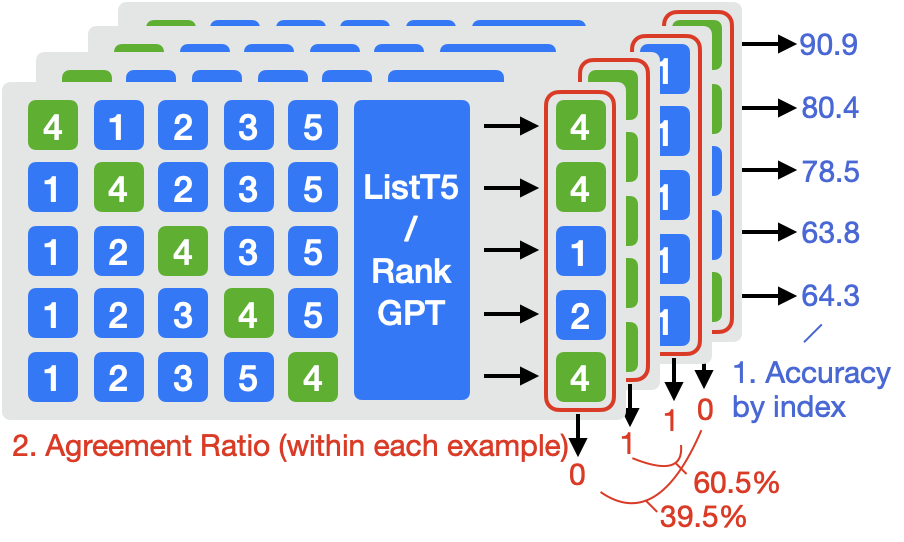}
    \caption{Measuring the robustness to positional bias by shuffling the index of the relevant passage. 4 is the gold (ground truth) relevant passage in the figure.}
    \label{fig:permutation_consistency}
}
\end{figure}
\paragraph{Background.} One of the biggest problems of lengthy input of zero-shot listwise rerankers is the lost in the middle problem~\cite{lostinthemiddle}. Recent studies show that LLMs exhibit strong positional bias to the information in the first and last parts of the input, and often fail to comprehend relevant information in the middle of a long input. A recent work also pointed out that this positional bias problem is also prevalent in applying LLMs to listwise reranking~\cite{foundinthemiddle}. 
The main problem arises from the different assignment of positional encoding. We believe using the FiD architecture effectively addresses this issue, as each passage is processed by the encoder with identical positional encoding. Consequently, the decoder \emph{cannot} exploit positional bias. To validate our statement, we conduct a comparative study to answer the following questions:

(1) How sensitive is the accuracy of each model to the position (location) of the positive passage? 

(2) How often does each model \textit{consistently} point to the same passage under position changes? 

\paragraph{Baseline Models.} We experiment with DuoT5 as representative for pairwise methods, and GPT3.5-turbo-1106 and GPT-4-0613 from OpenAI~\cite{openai}, the main model used for RankGPT~\cite{rankgpt}, as representative for listwise reranking methods. Note that we only consider the top-1 prediction of each model, so the choie of \textsc{ListT5} (r=1) and (r=2) does not affect the metric. Detailed explanation about setups with example prompts are in Appendix~\ref{ablation/permutation_consistency}.

\paragraph{Experiment and Data Setup.} We use the BM25 top 100 retrieved results from the FiQA-2018 and TREC-COVID test subset. 
For listwise ranking models, 
we randomly sample negative passages per every positive passage in every query to make groups of [1 positive, 4 negatives].\footnote{For TREC-COVID, one positive passage were sampled from each query since it has an average of 493 positive passages per each query.} Since pairwise models only take 2 passages for input, we additionally split the original group into 4 groups of [1 positive, 1 negative] for DuoT5. Similar with ~\citet{foundinthemiddle}, we measure the accuracy and agreement ratio after swapping the order of positive passages for each group (Fig.~\ref{fig:permutation_consistency}).
The results in Tab.~\ref{table/permutation_consistency_twocolumn} show that (1) our model suffers less from the positive index change with (2) a higher agreement ratio. 
Since each passage is distinguished by identifiers instead of positions, \textsc{ListT5} effectively overcomes the positional bias from long inputs by separately encoding each query-passage pair at the encoder and aggregating the output at the decoder. Still, the decoder can utilize and integrate the encoded information of all passages and effectively perform listwise reranking, achieving a better zero-shot performance than LLMs (Tab.~\ref{table/rankgpt_duot5}) or non-listwise counterparts (Tab.~\ref{table/main_table}).

\subsection{Impact of initial ordering}
\label{sec:results_initial_ordering}
\begin{table}[t]
    \resizebox{\columnwidth}{!}
    {
\begin{tabular}{@{}lcccccc@{}}
\toprule
\multicolumn{1}{l|}{\begin{tabular}[c]{@{}l@{}}Initial\\ ordering\end{tabular}} & \multicolumn{1}{c|}{DL19} & \multicolumn{1}{c|}{DL20} & \multicolumn{1}{c|}{\begin{tabular}[c]{@{}c@{}}TREC-\\COVID\end{tabular}} & \multicolumn{1}{c|}{\begin{tabular}[c]{@{}c@{}}TREC-\\NEWS\end{tabular}}& \multicolumn{1}{c|}{\begin{tabular}[c]{@{}c@{}}Touche\\-2020\end{tabular}} & Avg. \\ \midrule
\multicolumn{7}{l}{\textbf{ListT5-base (tournament sort, r=2)}} \\ \midrule
\multicolumn{1}{l|}{No shuffle} & \multicolumn{1}{c|}{71.8} & \multicolumn{1}{c|}{68.1} & \multicolumn{1}{c|}{78.3} & \multicolumn{1}{c|}{48.5} & \multicolumn{1}{c|}{33.4} & 60.0 \\
\multicolumn{1}{l|}{Shuffle} & \multicolumn{1}{c|}{71.2} & \multicolumn{1}{c|}{68.1} & \multicolumn{1}{c|}{77.2} & \multicolumn{1}{c|}{48.9} & \multicolumn{1}{c|}{32.8} & 59.6 \\
\multicolumn{1}{l|}{Perf. drop} & \multicolumn{1}{c|}{\textbf{}} & \multicolumn{1}{c|}{} & \multicolumn{1}{c|}{} & \multicolumn{1}{c|}{} & \multicolumn{1}{c|}{} & \multicolumn{1}{r}{\textbf{-0.4}} \\ \midrule
\multicolumn{7}{l}{ListT5-base (sliding windows, stride=3, iter=4)} \\ \midrule
\multicolumn{1}{l|}{No shuffle} & \multicolumn{1}{c|}{71.8} & \multicolumn{1}{c|}{67.7} & \multicolumn{1}{c|}{77.5} & \multicolumn{1}{c|}{50.0} & \multicolumn{1}{c|}{33.1} & 60.0 \\
\multicolumn{1}{l|}{Shuffle} & \multicolumn{1}{c|}{69.5} & \multicolumn{1}{c|}{65.5} & \multicolumn{1}{c|}{77.7} & \multicolumn{1}{c|}{49.2} & \multicolumn{1}{c|}{32.1} & 58.8 \\
\multicolumn{1}{l|}{Perf. drop} & \multicolumn{1}{c|}{} & \multicolumn{1}{c|}{} & \multicolumn{1}{c|}{} & \multicolumn{1}{c|}{} & \multicolumn{1}{c|}{} & \multicolumn{1}{r}{\textbf{-1.2}} \\ \midrule
\multicolumn{7}{l}{RankVicuna-7b (sliding windows)} \\ \midrule
\multicolumn{1}{l|}{No shuffle} & \multicolumn{1}{c|}{68.9} & \multicolumn{1}{c|}{66.1} & \multicolumn{1}{c|}{80.5} & \multicolumn{1}{c|}{46.9} & \multicolumn{1}{c|}{33.0} & 59.1 \\
\multicolumn{1}{l|}{Shuffle} & \multicolumn{1}{c|}{67.1} & \multicolumn{1}{c|}{64.6} & \multicolumn{1}{c|}{79.2} & \multicolumn{1}{c|}{45.3} & \multicolumn{1}{c|}{30.8} & 57.4 \\
\multicolumn{1}{l|}{Perf. drop} & \multicolumn{1}{c|}{} & \multicolumn{1}{c|}{} & \multicolumn{1}{c|}{} & \multicolumn{1}{c|}{} & \multicolumn{1}{c|}{} & \multicolumn{1}{r}{\textbf{-1.7}} \\ \midrule
\multicolumn{7}{l}{RankGPT-3.5 (sliding windows)} \\ \midrule
\multicolumn{1}{l|}{No shuffle} & \multicolumn{1}{c|}{68.4} & \multicolumn{1}{c|}{64.9} & \multicolumn{1}{c|}{72.6} & \multicolumn{1}{c|}{46.5} & \multicolumn{1}{c|}{38.2} & 58.1 \\
\multicolumn{1}{l|}{Shuffle} & \multicolumn{1}{c|}{62.5} & \multicolumn{1}{c|}{57.0} & \multicolumn{1}{c|}{66.1} & \multicolumn{1}{c|}{38.3} & \multicolumn{1}{c|}{22.8} & 49.3 \\
\multicolumn{1}{l|}{Perf. drop} & \multicolumn{1}{c|}{} & \multicolumn{1}{c|}{} & \multicolumn{1}{c|}{} & \multicolumn{1}{c|}{} & \multicolumn{1}{c|}{} & \multicolumn{1}{r}{\textbf{-8.8}} \\ \bottomrule
\end{tabular}
}
\caption{NDCG@10 results before and after randomly shuffling the initial top-100 ordering of BM25. Evaluation results for RankGPT-3.5 are run 3 times and averaged to compensate for the unstability of the API output. While RankGPT-3.5 suffers heavily after shuffling (-8.8 drop in performance), ListT5 with tournament sort is relatively robust to the initial order shuffling.}
\label{table/main_initial_ordering}
\end{table}
All listwise reranking methods, including the tournament sort and the sliding window approach, depends on the initial ordering returned by the first-stage retriever (e.g., BM25). This may leave a room for ordering sensitivity. 
To investigate this, we experiment on DL19, DL20, and the selected subset of BEIR, to measure the robustness of initial ordering. For each dataset, we shuffle the initial ordering of candidate passages on 3 different seeds (seed 0,1,2) and report the average score. The results from Table~\ref{table/main_initial_ordering} show that (1) ordering sensitivity is indeed prevalent in previous listwise reranking models, and that (2) using \textsc{ListT5} with FiD greatly reduces the sensitivity in general (in line with our findings in Section.~\ref{sec:results_lostinthemiddle}), and (3) When using \textsc{ListT5}, tournament sort is in general more robust to ordering change (-0.4 drop in average performance), compared to sliding windows (-1.2 drop). This means that \textsc{ListT5} with tournament sort can perform in a robust way even in conditions where the initial ordering is not very trustable, which is mostly the case for zero-shot retrieval tasks. 

\subsection{Ablation Study}
\begin{table}[t]
\small\addtolength{\tabcolsep}{-4pt}
\fontsize{6}{6}\selectfont
    \resizebox{\columnwidth}{!}
    {
\small
\begin{tabular}{@{}lccccc@{}}
\toprule
\multicolumn{1}{c|}{\multirow{2}{*}{Dataset}} & \multicolumn{1}{c|}{\multirow{2}{*}{\begin{tabular}[c]{@{}c@{}}Relevant\\ Discrimi-\\nation\end{tabular} }} & \multicolumn{2}{c|}{Relevant First}                      & \multicolumn{2}{c}{\begin{tabular}[c]{@{}c@{}}Relevant Last\\(ListT5)\end{tabular}}    \\ \cmidrule(lr){3-4} \cmidrule(lr){5-6} 
\multicolumn{1}{c|}{}                         & \multicolumn{1}{c|}{}                         & ($r$ = 1)        & \multicolumn{1}{c|}{($r$ = 2)}        & ($r = 1$)        & ($r = 2$)        \\ \midrule
\multicolumn{6}{l}{In-domain}                                                                                                                                                      \\ \midrule
\multicolumn{1}{l|}{MS MARCO}                 & \multicolumn{1}{c|}{40.3}                     & 40.8          & \multicolumn{1}{c|}{\textbf{40.9}} & 40.7          & 40.7          \\
\multicolumn{1}{l|}{TREC-DL19}                & \multicolumn{1}{c|}{\textbf{72.5}}            & 69.6          & \multicolumn{1}{c|}{70.8}          & 71.2          & 71.8          \\
\multicolumn{1}{l|}{TREC-DL20}                & \multicolumn{1}{c|}{67.3}                     & 67.0          & \multicolumn{1}{c|}{66.8}          & 67.3          & \textbf{68.1} \\ \midrule
\multicolumn{1}{l|}{\textbf{Avg (in-domain)}}           & \multicolumn{1}{c|}{60.0}                     & 59.1          & \multicolumn{1}{c|}{59.5}          & 59.7          & \textbf{60.2} \\ \midrule
\multicolumn{6}{l}{Out-domain (BEIR)}                                                                                                                                              \\ \midrule
\multicolumn{1}{l|}{TREC-COVID}               & \multicolumn{1}{c|}{74.0}                     & 74.9          & \multicolumn{1}{c|}{75.9}          & 76.7          & \textbf{78.3} \\
\multicolumn{1}{l|}{NFCorpus}                 & \multicolumn{1}{c|}{34.8}                     & 35.5          & \multicolumn{1}{c|}{35.6}          & 35.5          & \textbf{35.6} \\
\multicolumn{1}{l|}{BioASQ}                   & \multicolumn{1}{c|}{55.8}                     & 56.6          & \multicolumn{1}{c|}{56.6}          & \textbf{57.2} & 56.4          \\
\multicolumn{1}{l|}{NQ}                       & \multicolumn{1}{c|}{51.1}                     & 52.7          & \multicolumn{1}{c|}{52.9}          & 52.0          & \textbf{53.1} \\
\multicolumn{1}{l|}{HotpotQA}                 & \multicolumn{1}{c|}{70.9}                     & 72.5          & \multicolumn{1}{c|}{72.6}          & 72.1          & \textbf{72.6} \\
\multicolumn{1}{l|}{FiQA-2018}                & \multicolumn{1}{c|}{38.1}                     & 39.3          & \multicolumn{1}{c|}{39.0}          & 39.5          & \textbf{39.6} \\
\multicolumn{1}{l|}{Signal-1M (RT)}           & \multicolumn{1}{c|}{32.9}                     & 31.8          & \multicolumn{1}{c|}{31.7}          & 33.3          & \textbf{33.5} \\
\multicolumn{1}{l|}{TREC-NEWS}                & \multicolumn{1}{c|}{43.9}                     & 46.6          & \multicolumn{1}{c|}{47.3}          & 47.9          & \textbf{48.5} \\
\multicolumn{1}{l|}{Robust04}                 & \multicolumn{1}{c|}{49.8}                     & \textbf{52.3} & \multicolumn{1}{c|}{\textbf{52.3}} & 52.0          & 52.1          \\
\multicolumn{1}{l|}{Arguana}                  & \multicolumn{1}{c|}{26.1}                     & 32.8          & \multicolumn{1}{c|}{34.6}          & \textbf{49.7} & 48.9          \\
\multicolumn{1}{l|}{Touche-2020}              & \multicolumn{1}{c|}{\textbf{34.2}}            & 31.5          & \multicolumn{1}{c|}{31.3}          & \textbf{34.2} & 33.4          \\
\multicolumn{1}{l|}{CQADupStack}              & \multicolumn{1}{c|}{\textbf{38.8}}            & 38.3          & \multicolumn{1}{c|}{38.4}          & 38.4          & \textbf{38.8} \\
\multicolumn{1}{l|}{Quora}                    & \multicolumn{1}{c|}{81.9}                     & 84.4          & \multicolumn{1}{c|}{84.8}          & 86.1          & \textbf{86.4} \\
\multicolumn{1}{l|}{DBPedia}                  & \multicolumn{1}{c|}{42.4}                     & 43.4          & \multicolumn{1}{c|}{43.6}          & \textbf{43.9} & 43.7          \\
\multicolumn{1}{l|}{SCIDOCS}                  & \multicolumn{1}{c|}{16.3}                     & 17.3          & \multicolumn{1}{c|}{17.3}          & 17.2          & \textbf{17.6} \\
\multicolumn{1}{l|}{FEVER}                    & \multicolumn{1}{c|}{77.6}                     & 77.4          & \multicolumn{1}{c|}{77.7}          & 77.8          & \textbf{79.8} \\
\multicolumn{1}{l|}{Climate-FEVER}            & \multicolumn{1}{c|}{20.7}                     & 22.8          & \multicolumn{1}{c|}{23.0}          & 22.8          & \textbf{24.0} \\
\multicolumn{1}{l|}{SciFact}                  & \multicolumn{1}{c|}{73.0}                     & 74.1          & \multicolumn{1}{c|}{\textbf{74.2}} & 74.1          & 74.1          \\ \midrule
\multicolumn{1}{l|}{\textbf{Avg (BEIR)}}                & \multicolumn{1}{c|}{47.9}                     & 49.1          & \multicolumn{1}{c|}{49.4}          & 50.6          & \textbf{50.9} \\ \bottomrule
\end{tabular}

}
\caption{NDCG@10 results at in-domain and BEIR on varying the output format and generation order of passages. Generating from the least relevant passages shows better average performance on BEIR.}
\label{table/ablation_order}
\end{table}
\label{sec:results_design_choice}
\begin{figure}[!t]
{
\centering
    \includegraphics[width=\columnwidth]{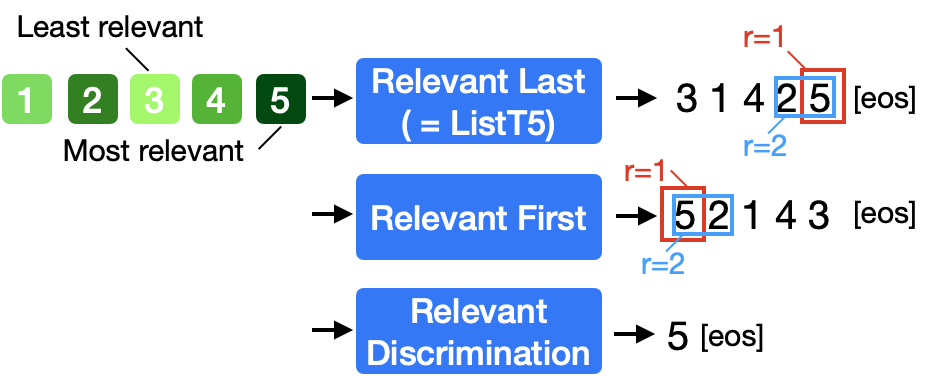}
    \caption{Variants of the output format of listwise ranking models while fixing the input.}
    \label{fig:fig_variants}
}
\end{figure}
To break down the role of each feature in our method, we perform experiments with different variants of our model (Fig.~\ref{fig:fig_variants}) in Tab.~\ref{table/ablation_order}. Additional ablation experiments, such as varying $m$, are in App.~\ref{appendix/ablation_topk}.

\paragraph{$\mathbf{r}$ = 1 vs $\mathbf{r}$ = 2.} Conducting inference by ($r$ = 2) improves the average BEIR performance by 0.3, compared with that of ($r$ = 1). 
Since this gain is also applicable to the Relevant First variants, which is the model trained to generate relevant index first, we hypothesize that $r$ = 2 serves as giving a second chance for the model to view passages with different candidates to make a decision with different viewpoints, where there are confusing passages (hard negatives), resulting in better ability to rank. 

\paragraph{Discrimination vs Sequential Sorting.} Given the same setup and dataset, we also train the model to output only relevant index, annotated as \textit{Relevant Discrimination}, similar to \citet{fidlight}. 
However, the Relevant Discrimination model shows the lowest performance amongst all model variants (47.9). 
We hypothesize that training the model to distinguish and order between negatives provides additional information for the model. While tasks that only generate positive indexes only learn the decision boundary between positive and negative contexts, learning to order negatives provides more informative signals and more accurate ranking results, since the model additionally learns to calibrate between any kind of passages. 

\paragraph{Relevant First vs Relevant Last (\textsc{ListT5}).}
We also compare models by changing the ordering of the output. Compared to \textsc{ListT5}, Relevant first shows a performance drop of 1-2 points in average on BEIR. As explained at Sec.~\ref{method/basic_operating_unit}, we conclude that generating in reversed order is important, as it effectively eliminates false negatives and acts as a reasoning chain. 

\section{Conclusion}
\label{sec:conclusion}
We proposed \textsc{ListT5}, a FiD-based listwise reranking model that jointly considers multiple passages as input and generates an ordered list of passages, in increasing order of relevance. \textsc{ListT5} outperforms others in BEIR, with a remarkable $+$1.3 gain of average NDCG@10 performance. \textsc{ListT5} is efficient, leveraging tournament sort to its advantage. 
Finally, we show that \textsc{ListT5} overcomes the lost-in-the-middle problem in LLMs, and are more robust regardless of the initial ordering of the passage. 

\newpage
\section{Limitations}
We believe that the efficiency of our models can be greatly optimized using simple ideas. For example, if we can implement early stopping at sequential decoding (assuming that the model outputs permutation of 5), we can reduce the decoding step by 80\% for \textsc{ListT5} ($r$=1) and 60\% For \textsc{ListT5} ($r$=2). There could also be other options, such as designing a more efficient way that uses the tournament sort with fewer number of forward passes, efficient way of encoder output caching, and extension to other langauges or other architectures.

We believe there are still space for optimization of \textsc{ListT5}, for example trying out $m$ other than 5 or 10, exploring with different learning rate or batch size on training, and so on. However, we were only able to experiment a few due to resource and time limitations.

Our results are mainly reported with BM25 Top-100, BM25 Top-1000, and COCO-DR Top-100 as first-stage retrieval modules. Also, our main results are based on the T5-base model and only a few were experimented with 3B models, due to the resource constraints. We hope to evaluate on extended setups for future work.

\section{Acknowledgements}
We greatly appreciate the members from the EXAONE lab from LG AI research for providing GPU resources to conduct experiments along with the helpful feedbacks. We would also like to thank the members from the SNU Language \& Data Intelligence lab for giving constructive feedbacks to improve the paper. We also thank the anonymous reviewers from ARR December and ARR February for pointing out important discussion points and additional experiments that help support our claim. Lastly, we would like to thank Jinwoo Kim from School of Computing, KAIST for helpful discussions.
\bibliography{anthology,custom}

\begin{thebibliography}{34}
\expandafter\ifx\csname natexlab\endcsname\relax\def\natexlab#1{#1}\fi

\bibitem[{Ai et~al.(2019)Ai, Wang, Bruch, Golbandi, Bendersky, and Najork}]{ai2019learning}
Qingyao Ai, Xuanhui Wang, Sebastian Bruch, Nadav Golbandi, Michael Bendersky, and Marc Najork. 2019.
\newblock Learning groupwise multivariate scoring functions using deep neural networks.
\newblock In \emph{Proceedings of the 2019 ACM SIGIR international conference on theory of information retrieval}, pages 85--92.

\bibitem[{Bajaj et~al.(2018)Bajaj, Campos, Craswell, Deng, Gao, Liu, Majumder, McNamara, Mitra, Nguyen, Rosenberg, Song, Stoica, Tiwary, and Wang}]{msmarco}
Payal Bajaj, Daniel Campos, Nick Craswell, Li~Deng, Jianfeng Gao, Xiaodong Liu, Rangan Majumder, Andrew McNamara, Bhaskar Mitra, Tri Nguyen, Mir Rosenberg, Xia Song, Alina Stoica, Saurabh Tiwary, and Tong Wang. 2018.
\newblock \href {http://arxiv.org/abs/1611.09268} {Ms marco: A human generated machine reading comprehension dataset}.

\bibitem[{Craswell et~al.(2021)Craswell, Mitra, Yilmaz, and Campos}]{trec20}
Nick Craswell, Bhaskar Mitra, Emine Yilmaz, and Daniel Campos. 2021.
\newblock \href {http://arxiv.org/abs/2102.07662} {Overview of the trec 2020 deep learning track}.

\bibitem[{Ferraretto et~al.(2023)Ferraretto, Laitz, Lotufo, and Nogueira}]{exaranker}
Fernando Ferraretto, Thiago Laitz, Roberto Lotufo, and Rodrigo Nogueira. 2023.
\newblock \href {http://arxiv.org/abs/2301.10521} {Exaranker: Explanation-augmented neural ranker}.

\bibitem[{Hofstätter et~al.(2022)Hofstätter, Chen, Raman, and Zamani}]{fidlight}
Sebastian Hofstätter, Jiecao Chen, Karthik Raman, and Hamed Zamani. 2022.
\newblock \href {http://arxiv.org/abs/2209.14290} {Fid-light: Efficient and effective retrieval-augmented text generation}.

\bibitem[{Izacard and Grave(2021)}]{fid}
Gautier Izacard and Edouard Grave. 2021.
\newblock \href {http://arxiv.org/abs/2007.01282} {Leveraging passage retrieval with generative models for open domain question answering}.

\bibitem[{J\"{a}rvelin and Kek\"{a}l\"{a}inen(2002)}]{ndcg}
Kalervo J\"{a}rvelin and Jaana Kek\"{a}l\"{a}inen. 2002.
\newblock \href {https://doi.org/10.1145/582415.582418} {Cumulated gain-based evaluation of ir techniques}.
\newblock \emph{ACM Trans. Inf. Syst.}, 20(4):422–446.

\bibitem[{Karpukhin et~al.(2020)Karpukhin, Oguz, Min, Lewis, Wu, Edunov, Chen, and Yih}]{dpr}
Vladimir Karpukhin, Barlas Oguz, Sewon Min, Patrick Lewis, Ledell Wu, Sergey Edunov, Danqi Chen, and Wen-tau Yih. 2020.
\newblock \href {https://doi.org/10.18653/v1/2020.emnlp-main.550} {Dense passage retrieval for open-domain question answering}.
\newblock In \emph{Proceedings of the 2020 Conference on Empirical Methods in Natural Language Processing (EMNLP)}, pages 6769--6781, Online. Association for Computational Linguistics.

\bibitem[{Lakhotia et~al.(2020)Lakhotia, Paranjape, Ghoshal, tau Yih, Mehdad, and Iyer}]{fidex}
Kushal Lakhotia, Bhargavi Paranjape, Asish Ghoshal, Wen tau Yih, Yashar Mehdad, and Srinivasan Iyer. 2020.
\newblock \href {http://arxiv.org/abs/2012.15482} {Fid-ex: Improving sequence-to-sequence models for extractive rationale generation}.

\bibitem[{Liu et~al.(2023)Liu, Lin, Hewitt, Paranjape, Bevilacqua, Petroni, and Liang}]{lostinthemiddle}
Nelson~F. Liu, Kevin Lin, John Hewitt, Ashwin Paranjape, Michele Bevilacqua, Fabio Petroni, and Percy Liang. 2023.
\newblock \href {http://arxiv.org/abs/2307.03172} {Lost in the middle: How language models use long contexts}.

\bibitem[{Ma et~al.(2023)Ma, Zhang, Pradeep, and Lin}]{listwisereranking}
Xueguang Ma, Xinyu Zhang, Ronak Pradeep, and Jimmy Lin. 2023.
\newblock \href {http://arxiv.org/abs/2305.02156} {Zero-shot listwise document reranking with a large language model}.

\bibitem[{McLuckie and Barber(1986)}]{tournamentsort}
Keith McLuckie and Angus Barber. 1986.
\newblock \href {https://doi.org/10.1007/978-1-349-08147-9_5} {\emph{Tournament Sort}}, pages 68--86. Macmillan Education UK, London.

\bibitem[{Murphy et~al.(2019)Murphy, Srinivasan, Rao, and Ribeiro}]{pairwise_murphy}
Ryan Murphy, Balasubramaniam Srinivasan, Vinayak Rao, and Bruno Ribeiro. 2019.
\newblock \href {https://proceedings.mlr.press/v97/murphy19a.html} {Relational pooling for graph representations}.
\newblock In \emph{Proceedings of the 36th International Conference on Machine Learning}, volume~97 of \emph{Proceedings of Machine Learning Research}, pages 4663--4673. PMLR.

\bibitem[{Ni et~al.(2022)Ni, Qu, Lu, Dai, Hernandez~Abrego, Ma, Zhao, Luan, Hall, Chang, and Yang}]{gtr}
Jianmo Ni, Chen Qu, Jing Lu, Zhuyun Dai, Gustavo Hernandez~Abrego, Ji~Ma, Vincent Zhao, Yi~Luan, Keith Hall, Ming-Wei Chang, and Yinfei Yang. 2022.
\newblock \href {https://doi.org/10.18653/v1/2022.emnlp-main.669} {Large dual encoders are generalizable retrievers}.
\newblock In \emph{Proceedings of the 2022 Conference on Empirical Methods in Natural Language Processing}, pages 9844--9855, Abu Dhabi, United Arab Emirates. Association for Computational Linguistics.

\bibitem[{Nogueira et~al.(2020)Nogueira, Jiang, Pradeep, and Lin}]{monot5}
Rodrigo Nogueira, Zhiying Jiang, Ronak Pradeep, and Jimmy Lin. 2020.
\newblock \href {https://doi.org/10.18653/v1/2020.findings-emnlp.63} {Document ranking with a pretrained sequence-to-sequence model}.
\newblock In \emph{Findings of the Association for Computational Linguistics: EMNLP 2020}, pages 708--718, Online. Association for Computational Linguistics.

\bibitem[{OpenAI(2022)}]{openai}
OpenAI. 2022.
\newblock Introducing chatgpt.
\newblock \url{https://openai.com/blog/chatgpt}.

\bibitem[{Pradeep et~al.(2021)Pradeep, Nogueira, and Lin}]{duot5}
Ronak Pradeep, Rodrigo Nogueira, and Jimmy Lin. 2021.
\newblock \href {http://arxiv.org/abs/2101.05667} {The expando-mono-duo design pattern for text ranking with pretrained sequence-to-sequence models}.

\bibitem[{Pradeep et~al.(2023{\natexlab{a}})Pradeep, Sharifymoghaddam, and Lin}]{rankvicuna}
Ronak Pradeep, Sahel Sharifymoghaddam, and Jimmy Lin. 2023{\natexlab{a}}.
\newblock \href {http://arxiv.org/abs/2309.15088} {Rankvicuna: Zero-shot listwise document reranking with open-source large language models}.

\bibitem[{Pradeep et~al.(2023{\natexlab{b}})Pradeep, Sharifymoghaddam, and Lin}]{rankzephyr}
Ronak Pradeep, Sahel Sharifymoghaddam, and Jimmy Lin. 2023{\natexlab{b}}.
\newblock \href {http://arxiv.org/abs/2312.02724} {Rankzephyr: Effective and robust zero-shot listwise reranking is a breeze!}

\bibitem[{Qin et~al.(2023)Qin, Jagerman, Hui, Zhuang, Wu, Shen, Liu, Liu, Metzler, Wang, and Bendersky}]{llms-effective-rankers}
Zhen Qin, Rolf Jagerman, Kai Hui, Honglei Zhuang, Junru Wu, Jiaming Shen, Tianqi Liu, Jialu Liu, Donald Metzler, Xuanhui Wang, and Michael Bendersky. 2023.
\newblock \href {http://arxiv.org/abs/2306.17563} {Large language models are effective text rankers with pairwise ranking prompting}.

\bibitem[{Raffel et~al.(2020)Raffel, Shazeer, Roberts, Lee, Narang, Matena, Zhou, Li, and Liu}]{t5}
Colin Raffel, Noam Shazeer, Adam Roberts, Katherine Lee, Sharan Narang, Michael Matena, Yanqi Zhou, Wei Li, and Peter~J. Liu. 2020.
\newblock \href {http://jmlr.org/papers/v21/20-074.html} {Exploring the limits of transfer learning with a unified text-to-text transformer}.
\newblock \emph{Journal of Machine Learning Research}, 21(140):1--67.

\bibitem[{Reimers and Gurevych(2019)}]{sentencebert}
Nils Reimers and Iryna Gurevych. 2019.
\newblock \href {http://arxiv.org/abs/1908.10084} {Sentence-bert: Sentence embeddings using siamese bert-networks}.

\bibitem[{Ren et~al.(2018)Ren, Liu, and Shroff}]{ren2018pac}
Wenbo Ren, Jia Liu, and Ness~B. Shroff. 2018.
\newblock \href {http://arxiv.org/abs/1806.02970} {Pac ranking from pairwise and listwise queries: Lower bounds and upper bounds}.

\bibitem[{Robertson and Zaragoza(2009)}]{bm25}
Stephen Robertson and Hugo Zaragoza. 2009.
\newblock \href {https://doi.org/10.1561/1500000019} {The probabilistic relevance framework: Bm25 and beyond}.
\newblock \emph{Foundations and Trends in Information Retrieval}, 3(4):333–389.

\bibitem[{Rosa et~al.(2022)Rosa, Bonifacio, Jeronymo, Abonizio, Fadaee, Lotufo, and Nogueira}]{indefense}
Guilherme Rosa, Luiz Bonifacio, Vitor Jeronymo, Hugo Abonizio, Marzieh Fadaee, Roberto Lotufo, and Rodrigo Nogueira. 2022.
\newblock \href {http://arxiv.org/abs/2212.06121} {In defense of cross-encoders for zero-shot retrieval}.

\bibitem[{Sun et~al.(2023{\natexlab{a}})Sun, Chen, Ma, Yan, Wang, Ren, Chen, Yin, and Ren}]{distillrankgpt}
Weiwei Sun, Zheng Chen, Xinyu Ma, Lingyong Yan, Shuaiqiang Wang, Pengjie Ren, Zhumin Chen, Dawei Yin, and Zhaochun Ren. 2023{\natexlab{a}}.
\newblock \href {http://arxiv.org/abs/2311.01555} {Instruction distillation makes large language models efficient zero-shot rankers}.

\bibitem[{Sun et~al.(2023{\natexlab{b}})Sun, Yan, Ma, Wang, Ren, Chen, Yin, and Ren}]{rankgpt}
Weiwei Sun, Lingyong Yan, Xinyu Ma, Shuaiqiang Wang, Pengjie Ren, Zhumin Chen, Dawei Yin, and Zhaochun Ren. 2023{\natexlab{b}}.
\newblock \href {http://arxiv.org/abs/2304.09542} {Is chatgpt good at search? investigating large language models as re-ranking agents}.

\bibitem[{Tang et~al.(2023)Tang, Zhang, Ma, Lin, and Ture}]{foundinthemiddle}
Raphael Tang, Xinyu Zhang, Xueguang Ma, Jimmy Lin, and Ferhan Ture. 2023.
\newblock \href {http://arxiv.org/abs/2310.07712} {Found in the middle: Permutation self-consistency improves listwise ranking in large language models}.

\bibitem[{Thakur et~al.(2021)Thakur, Reimers, Rücklé, Srivastava, and Gurevych}]{beir}
Nandan Thakur, Nils Reimers, Andreas Rücklé, Abhishek Srivastava, and Iryna Gurevych. 2021.
\newblock \href {http://arxiv.org/abs/2104.08663} {Beir: A heterogenous benchmark for zero-shot evaluation of information retrieval models}.

\bibitem[{Wang et~al.(2024)Wang, Yang, Huang, Yang, Majumder, and Wei}]{wang2024large}
Liang Wang, Nan Yang, Xiaolong Huang, Linjun Yang, Rangan Majumder, and Furu Wei. 2024.
\newblock \href {http://arxiv.org/abs/2310.14587} {Large search model: Redefining search stack in the era of llms}.

\bibitem[{Xian et~al.(2023)Xian, Zhuang, Qin, Zamani, Lu, Ma, Hui, Zhao, Wang, and Bendersky}]{xian2023learning}
Ruicheng Xian, Honglei Zhuang, Zhen Qin, Hamed Zamani, Jing Lu, Ji~Ma, Kai Hui, Han Zhao, Xuanhui Wang, and Michael Bendersky. 2023.
\newblock \href {http://arxiv.org/abs/2212.10764} {Learning list-level domain-invariant representations for ranking}.

\bibitem[{Yarotsky(2018)}]{pairwise_yarotsky}
Dmitry Yarotsky. 2018.
\newblock \href {http://arxiv.org/abs/1804.10306} {Universal approximations of invariant maps by neural networks}.

\bibitem[{Yu et~al.(2022)Yu, Xiong, Sun, Zhang, and Overwijk}]{cocodr}
Yue Yu, Chenyan Xiong, Si~Sun, Chao Zhang, and Arnold Overwijk. 2022.
\newblock \href {http://arxiv.org/abs/2210.15212} {Coco-dr: Combating distribution shifts in zero-shot dense retrieval with contrastive and distributionally robust learning}.

\bibitem[{Zhuang et~al.(2022)Zhuang, Qin, Jagerman, Hui, Ma, Lu, Ni, Wang, and Bendersky}]{rankt5}
Honglei Zhuang, Zhen Qin, Rolf Jagerman, Kai Hui, Ji~Ma, Jing Lu, Jianmo Ni, Xuanhui Wang, and Michael Bendersky. 2022.
\newblock \href {http://arxiv.org/abs/2210.10634} {Rankt5: Fine-tuning t5 for text ranking with ranking losses}.

\end{thebibliography}
\bibliographystyle{acl_natbib}
\newpage
\appendix
\newpage

\section{Related Work - FiD in a Similar Setup}
\citet{fidlight} assigned index numbers to each query and passage pair, directing an FiD model to generate relevant passage indexes—an idea initially proposed by \citet{fidex}. 
While their approach is similar to ours, it diverges in that they only generate a single positive index, not the sorted list of all indices.

\section{Design Choice of \textsc{ListT5}}
\subsection{Impact of the number of passages $m$ \textsc{ListT5} sees at once}
\label{appendix/ablation_topk}
\begin{table}
\centering
    \resizebox{0.9\columnwidth}{!}
    {
\renewcommand{\arraystretch}{1.15}
\begin{tabular}{@{}l|cccccc@{}}
\toprule
         & \multicolumn{4}{c}{$m$ = 5 ( = \textsc{ListT5})}                                                                               & \multicolumn{2}{c}{$m$ = 10} \\ \midrule
         & \begin{tabular}[c]{@{}c@{}}1/5\\ (r=1)\end{tabular} & \begin{tabular}[c]{@{}c@{}}2/5\\(r=2)\end{tabular} & \begin{tabular}[c]{@{}c@{}}3/5\\ (r=3)\end{tabular} & \begin{tabular}[c]{@{}c@{}}4/5\\(r=4)\end{tabular} & \begin{tabular}[c]{@{}c@{}}1/10\\ (r=1)\end{tabular} & \begin{tabular}[c]{@{}c@{}}4/10\\ (r=4)\end{tabular} \\ \midrule
MS MARCO & \textbf{40.7}                                          & \textbf{40.7}                                            & -        & -         & 40.5         & \textbf{40.7}         \\
\multicolumn{1}{c|}{+ Top-1000}      & 44.7          & 44.9          & - & - & 44.6 & \textbf{45.0} \\
TREC-DL19      & 71.2          & \textbf{71.8} & - & -         & 70.1 & 70.5         \\
TREC-DL20      & 67.3          & \textbf{68.1} & - & -          & 66.9 & 67.2          \\ \midrule
TREC-COVID     & 76.7          & \textbf{78.3} & 78.1 & \textbf{78.3}          & 76.2 & 77.9          \\
NFCorpus       & 35.5          & 35.6          & 35.3 & 35.6          & 36.2 & \textbf{36.6} \\
BioASQ         & \textbf{57.2} & 56.4          & 55.9 & 55.8          & 55.4 & 56.4          \\
NQ             & 52.0          & \textbf{53.1} & 52.2 & 52.2          & 51.5 & 52.5          \\
HotpotQA       & 72.1          & \textbf{72.6} & 71.6 & 71.6          & 71.4 & 71.9          \\
FiQA-2018      & 39.5          & \textbf{39.6} & \textbf{39.6} & 39.5          & 39.0 & 38.9          \\
Signal-1M (RT) & 33.3          & \textbf{33.5} & 33.3 & 33.2          & 31.7 & 32.0          \\
TREC-NEWS      & 47.9          & 48.5 & \textbf{49.2} & 48.6          & 47.1 & 47.8          \\
Robust04       & 52.0          & 52.1          & 51.6 & 51.7          & 52.2 & \textbf{53.1} \\
Arguana        & 49.7  & 48.9          & 49.4 & \textbf{49.9}          & 38.6 & 46.6          \\
Touche-2020    & \textbf{34.2} & 33.4          & 33.8 & \textbf{34.2}          & 32.4 & 32.7          \\
CQADupStack    & 38.4          & \textbf{38.8} & 38.5 & 38.6          & 38.2 & 28.8          \\
Quora          & 86.1          & 86.4          & 86.0 & 86.1          & 85.5 & \textbf{86.8} \\
DBPedia        & \textbf{43.9}          & 43.7 & 43.3 & 43.1          & 42.7 & 43.6          \\
SCIDOCS        & 17.2          & 17.6          & 17.4 & 17.5          & 17.2 & \textbf{18.0} \\
FEVER          & 77.8          & \textbf{79.8} & 78.6 & 78.4          & 76.7 & 79.1          \\
Climate-FEVER  & 22.8          & \textbf{24.0} & 23.4 & 23.3          & 22.7 & 23.9          \\
SciFact        & 74.1          & 74.1          & 73.8 & 74.3          & 73.4 & \textbf{74.2} \\ \midrule
Avg(In-domain) & 56.0          & \textbf{56.4} & - & -          & 55.5 & 55.9          \\
Avg(BeIR)      & 50.6          & \textbf{50.9} & 50.6 & 50.7          & 49.3 & 50.0          \\ \bottomrule
\end{tabular}
}
\caption{NDCG@10 performance results for in-domain \& BEIR with \textsc{ListT5-base}, on differing \textbf{$m$}. (Sec.\ref{appendix/ablation_topk}.) We compare evaluation results that are trained on seeing 5 / 10 passages at once. Like the original \textsc{ListT5}, both models generate relevant indexes at the last.}
\label{table/appendix_top10}
\end{table}

This aligns with Sec.~\ref{sec:results_design_choice} from the main paper.
In addition to varying the output format, we try to increase the number of passages $m$ the model sees at once, with variants on $r$ as well (See Sec.~\ref{method/basic_operating_unit} for basic notation). The results from Tab.~\ref{table/appendix_top10}  show that setting $m$ as 10 is not as effective as setting $m$ as 5.
We conjecture that there is a trade-off between the (1) benefits from seeing more negatives at once, and (2) the complexity of the ordering task itself. If we increase $m$, the model can get more information, but at the same time, the task becomes very hard, and the boundary of negative order becomes very weak (and not very informative). Therefore, in the case of T5-base models, we find that 5 performed better than 10 passages.
Compared to the original FiD for QA~\cite{fid} where they use large $m$ (=100), ListT5 works best with smaller $m$. We conjecture the reason for this is that sorting task has more complexity than the original task that only use the input as knowledge source, leading to increased effectiveness when $m$ is smaller. Where the use case of the original FiD are to find relevant information among inputs, \textsc{ListT5} needs to order between different inputs, a task that needs to analyze the relative relevancy between \textit{all} given inputs. However, we believe the optimal number can be improved (or can be different) for models with different architectures or with larger parametric size.

\subsection{Impact of the negative selection model}
\label{appendix/gtr}
\begin{table}[t]
\centering
    \resizebox{\columnwidth}{!}
    {
\begin{tabular}{@{}l|c|cccc@{}}
\toprule
Model & RankT5 & \multicolumn{4}{c}{ListT5} \\ \midrule
Training data & GTR & \multicolumn{1}{c|}{COCO-DR} & \multicolumn{3}{c}{GTR} \\ \cmidrule(lr){1-1} 
 \cmidrule(lr){2-2} \cmidrule(lr){3-3} \cmidrule(lr){4-6}
Learning Rate & 1.00E-04 & \multicolumn{1}{c|}{1.00E-04} & 1.00E-04 & 1.00E-04 & 1.00E-05 \\
Steps & - & \multicolumn{1}{c|}{20k} & 20k & 10k & 30k \\ \midrule
TREC-COVID & 77.7 & \multicolumn{1}{c|}{78.3} & 77.3 & 78.6 & 77.9 \\
NFCorpus & 35.1 & \multicolumn{1}{c|}{35.6} & 35.4 & 36.2 & 35.9 \\
BioASQ & 58.2 & \multicolumn{1}{c|}{56.4} & 54.9 & 55.1 & 56.8 \\
NQ & 53.2 & \multicolumn{1}{c|}{53.1} & 52.7 & 52.8 & 53.2 \\
HotpotQA & 72.8 & \multicolumn{1}{c|}{72.6} & 72.1 & 71.9 & 72.1 \\
FiQA-2018 & 39.2 & \multicolumn{1}{c|}{39.6} & 39.1 & 39.9 & 39.4 \\
Signal-1M (RT) & 30.8 & \multicolumn{1}{c|}{33.5} & 34.1 & 32.9 & 30.9 \\
TREC-NEWS & 45.4 & \multicolumn{1}{c|}{48.5} & 47.6 & 48.0 & 48.3 \\
Robust04 & 54.3 & \multicolumn{1}{c|}{52.1} & 52.9 & 52.7 & 53.6 \\
Arguana & 35.5 & \multicolumn{1}{c|}{48.9} & 43.3 & 43.6 & 43.7 \\
Touche-2020 & 37.1 & \multicolumn{1}{c|}{33.4} & 31.5 & 32.7 & 32.5 \\
CQADupStack & 37.0 & \multicolumn{1}{c|}{38.8} & 38.6 & 38.5 & 38.8 \\
Quora & 83.3 & \multicolumn{1}{c|}{86.4} & 86.0 & 83.9 & 84.4 \\
DBPedia & 43.7 & \multicolumn{1}{c|}{43.7} & 43.8 & 43.6 & 44.5 \\
SCIDOCS & 16.8 & \multicolumn{1}{c|}{17.6} & 17.1 & 17.1 & 17.6 \\
FEVER & 77.6 & \multicolumn{1}{c|}{79.8} & 78.9 & 79.4 & 79.7 \\
Climate-FEVER & 21.2 & \multicolumn{1}{c|}{24.0} & 24.0 & 24.8 & 24.6 \\
SciFact & 73.5 & \multicolumn{1}{c|}{74.1} & 74.0 & 73.1 & 74.4 \\ \midrule
Avg. & 49.6 & \multicolumn{1}{c|}{\textbf{50.9}} & 50.2 & 50.3 & \textbf{50.5} \\ \bottomrule
\end{tabular}

}
\caption{Comparison of ListT5 models trained with GTR and COCO-DR. For ListT5, the reported scores are evaluated using the (r=2) variant. (Sec.~\ref{appendix/gtr})}
\label{table/appendix_gtr}
\end{table}

In our main experiment, we use COCO-DR to select and align negatives from MS MARCO, when constructing the training data. To analyze the impact of using COCO-DR as negatives, we also experiment with training ListT5 using negatives sampled from GTR,\footnote{\url{https://huggingface.co/sentence-transformers/gtr-t5-xl}}  since RankT5~\cite{rankt5} used GTR to retrieve top 1000 passage from each query and conduct random sampling to select negative passages. (Note that RankT5 samples 35 negative passages per one query from the top 1000 retrieval results, while ListT5 only sees 4 negatives at once.) The results are at Tab.~\ref{table/appendix_gtr}. First, we train the T5-base model with the same hyperparameter setup for those used to train the original \textsc{ListT5}-COCO-DR model. \footnote{With learning rate of 1e-04, linear learning rate scheduler with 10\% warmup, effective batch size 256, maximum input length 230, and maximum output length of 7, and train the T5-base model for 20k steps.} Then, we conduct hyperparameter search on the learning rate and number of steps for the GTR model, and found that training the model with learning rate of 1e-05 for 30k steps performs best (average 50.5). The results indicate that while \textsc{ListT5} trained with different negatives can impact performance, the performance gap is not significant, and the ListT5 model based on both COCO-DR and GTR still exhibit better performance than RankT5.

\section{Details about Our Inference Framework}
\label{appendix/inference}
This aligns with Sec.~\ref{method/extension_torunament_sort} from the main paper.
\subsection{Background: Tournament Sort}
\label{appendix/tournament_tree}
Tournament sort is a variation of heap sort, inspired by the concept from sports tournaments, where its objective is to identify not just the best player, but also the $k$-best, with the minimum number of games. While naive selection sort takes $\mathcal{O}(n)$ operations to iteratively select the best element, \textit{tournament sort} 
leverage its tournament tree structure for efficient sorting, resulting in needing only $\mathcal{O}(\log n)$ operations after the initial building of the tree in $\mathcal{O}(n)$

\subsection{Modification to Accept \textsc{ListT5} ($\mathbf{r}$ = 2)}
\begin{figure*}[!t]
{
\centering
    \includegraphics[width=\textwidth]{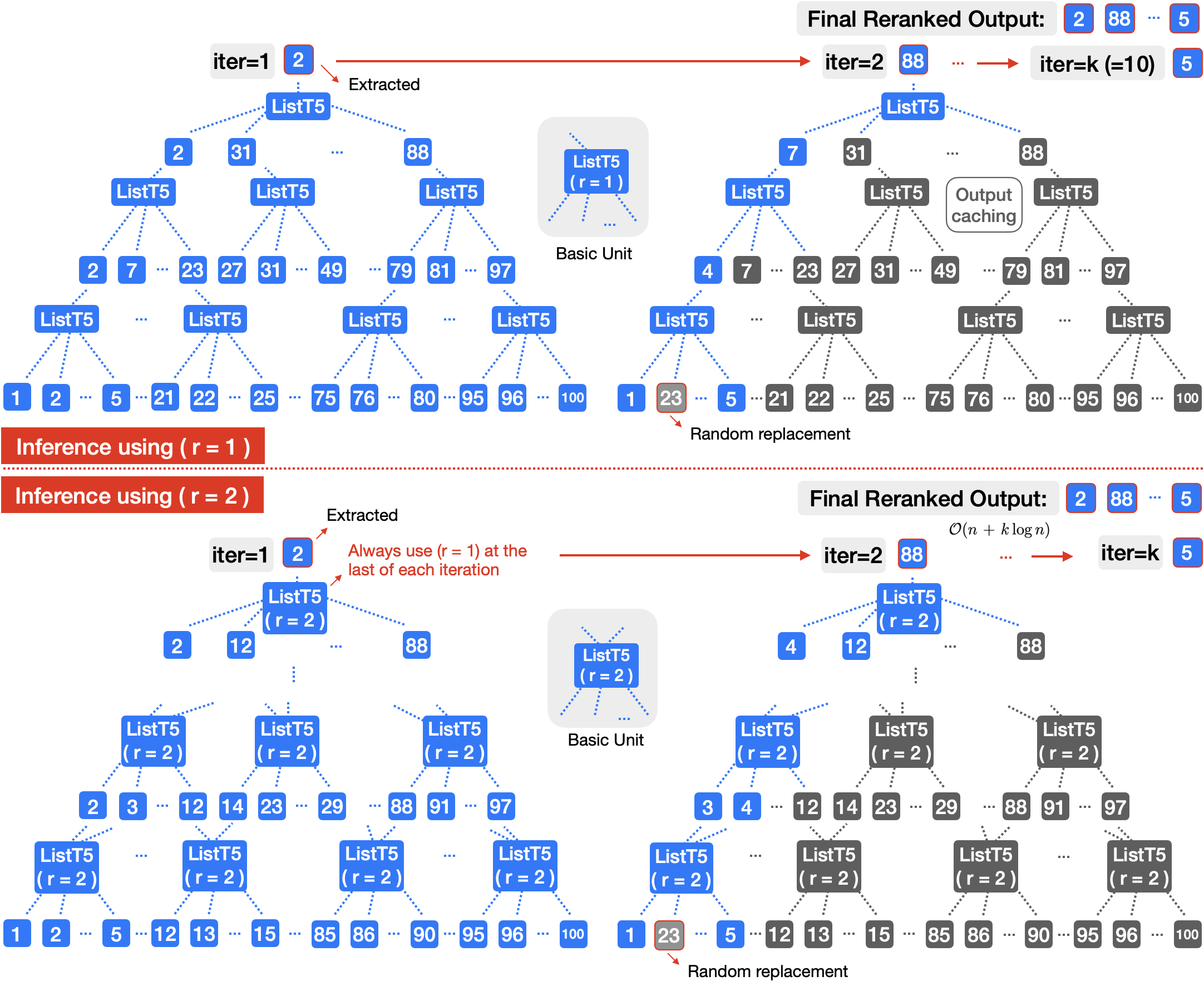}
    \caption{Illustration of our inference scenario. We show and compare the process with ($r$=1) and ($r$=2).}
    \label{fig:fig_appendix_inference}
}
\end{figure*}

Fig.~\ref{fig:fig_appendix_inference} illustrates and compare the inference scenario for \textsc{ListT5} ($r$ = 1) and \textsc{ListT5} ($r$ = 2). Since the ($r$=2) variant outputs \textit{2} different passage indices, the total number of candidate passages we need to compute is multipled by 2 for the next level of a tree. For example, if we use \textsc{ListT5} ($r$ = 1) with $m$ = 5 at the initial iteration, after computation of the bottom-level, we would get a total of 20 (100 / 5) filtered candidates which will go to the next round for comparison. In contrast, if we use \textsc{ListT5} ($r$ = 2) with $m$ = 5, we would get a total of 40 (100/5 $*$ 2) candidates to compute at the next round. However, it is important to note that we only use the inference method of ($r$ = 1) for the root computation, since we always outputs the top-1 candidates for every iteration.

\subsection{Details on Random Replacement}
\label{appendix_sec_random_replacement}
As explained as the 3rd and 4th modification at the main paper Sec.~\ref{method/extension_torunament_sort}, we explain in detail the random assignment process here.
Unlike the original tournament sort where they mark the final selected values to infinity, it is not an option for us since we cannot define a passage that has infinity scores of relevance. Therefore, we choose to randomly fill out the place, by the following rules;
\paragraph{Random assignment of top-1 selected passage}
After one iteration, one index is selected as most relevant out of $n$ candidate passages. Then, we add that index into the global exclude pool and remove the index from the consideration pool. After that, using the initial ordering of the passage by the first-stage retrieval, we add $+$21 to the original selected index. (For example, if the selected index was 8, we replace that passage with 29. We just set it to a larger value than $m$=5 to avoid meeting duplicates, but it can be anything) If the index is already in the global exclude pool \textit{or} have duplicates inside the input window $m$, we consider the next passage by adding 1 until the conditions are met.

\paragraph{Random assignment of dummy values to fill $\mathbf{m}$}
\begin{table}[t!]
\centering
    \resizebox{\columnwidth}{!}
    {
\begin{tabular}{@{}l|cccccccc@{}}
\toprule
 & \multicolumn{8}{c}{Random replace index number:} \\ \midrule
Dataset & \begin{tabular}[c]{@{}c@{}}21\\ (orig.)\end{tabular} & 20 & 19 & 18 & 17 & 16 & 15 & 6 \\ \midrule
signal & 33.5 & 33.3 & 33.2 & 33.1 & 33.3 & 33.3 & 33.3 & 33.5 \\
trec-covid & 78.3 & 78.2 & 78.2 & 78.4 & 78.1 & 78.1 & 78.3 & 78.3 \\
news & 48.5 & 48.4 & 48.4 & 48.4 & 48.3 & 48.4 & 48.4 & 48.4 \\
scifact & 74.1 & 74.0 & 74.0 & 74.0 & 74.0 & 74.1 & 74.2 & 74.0 \\
nfcorpus & 35.6 & 35.7 & 35.7 & 35.7 & 35.7 & 35.7 & 35.7 & 35.8 \\
fiqa & 39.6 & 39.6 & 39.7 & 39.7 & 39.6 & 39.6 & 39.6 & 39.7 \\
bioasq & 56.4 & 56.6 & 56.5 & 56.5 & 56.4 & 56.5 & 56.5 & 56.5 \\
touche & 33.4 & 33.7 & 33.4 & 33.5 & 33.6 & 33.8 & 33.8 & 33.6 \\
dbpedia & 43.7 & 43.6 & 43.8 & 43.7 & 43.8 & 43.7 & 43.7 & 43.8 \\
robust04 & 52.1 & 52.1 & 52.0 & 52.1 & 52.0 & 52.1 & 52.1 & 51.9 \\
scidocs & 17.6 & 17.6 & 17.6 & 17.5 & 17.6 & 17.5 & 17.6 & 17.5 \\
arguana & 48.9 & 49.0 & 48.8 & 48.8 & 48.9 & 49.0 & 48.9 & 48.9 \\
climate-fever & 24.0 & 24.0 & 24.0 & 24.0 & 24.0 & 24.0 & 24.0 & 24.0 \\
dl19 & 71.8 & 71.9 & 71.8 & 71.7 & 71.7 & 71.8 & 71.7 & 71.8 \\
dl20 & 68.1 & 68.2 & 68.1 & 68.1 & 68.1 & 68.1 & 68.3 & 68.5 \\
Average & \textbf{48.4} & \textbf{48.4} & \textbf{48.3} & \textbf{48.3} & \textbf{48.3} & \textbf{48.4} & \textbf{48.4} & \textbf{48.4} \\ \bottomrule
\end{tabular}
}
\caption{NDCG@10 results on differing the index of random assignment, with explanations on Section~\ref{appendix_sec_random_replacement}.}
\label{table/appendix_random_replace_number}
\end{table}
The basic unit of \textsc{ListT5} always accepts exactly $m$ passages as the input. However, there are cases when the leftover passages are not a multiple of $m$. For example, as we proceed to the upper level on the tournament tree, there will be a total of 100 $\mapsto$ 20 $\mapsto$ 4 passages left if we rank 100 passages using \textsc{ListT5} ($r$=1) with $m$=5. In this case, we would need one more dummy passage to make 4 candidate passages to 5. For reproducibility, we select a dummy passage in a deterministic way by considering the first passage from the initial candidate passage pool. If that passage is already selected as top-$k$ indices during the previous round \textit{or} we have duplicates inside the input window $m$, we proceed to the next index until the conditions are met. To show that the assignment of the random index is not statistically significant, we have conducted additional experiments on the subset of BEIR with differing the random replace index number. The results are presented at Table.~\ref{table/appendix_random_replace_number}. From the table, we can see that the output is robust to the assignment of random replace index number.

\paragraph{Edge Cases.}
There are some cases, especially for the case of NFCorpus, that the initial retrieval module doesn't give exactly $n$ passages as initial candidate passages. Sometimes, they give $n$ that is smaller than $m$ (e.g., 1,2,3,4). For this case, we exceptionally allow duplicates within the input window $m$, run one basic unit of \textsc{ListT5} with (r=$n$) (We remove duplicate passage indexes so that the output ordering reduces to $n$). That is, we only run the inference once, and use the output ordering as it is. Also, if $k$ is bigger than $n$-$m$, we inevitably meet the case where the leftover (unselected) passage number becomes smaller than $m$. (For example, ranking the 96th candidate passage for $n$=100 with $m$=5). Similar to the previous case, we also allow duplicates for this case and run the basic unit once to order the leftover indices.

\subsection{Example Scenario using Output Caching}
For example, if we consider reranking top 10 passages from 100 candidate passages ($n$ = 100, $k$ = 10), without output caching results in 250 forward passes in total; since we would need ((100/5) + (20/5) + 1) = 25 forward passes for each iteration, the total will be 25 * 10 = 250. With output caching, after the initial iteration, we only need to do one computation for each level of a tree, resulting in (1+1+1) = 3 forward passes. Summing up results in 25 + 3*9 = 54 forward passes, almost reducing \textbf{80\%} of the total number of inferences.

\section{Additional Results.}
\subsection{Results on COCO-DR as first-stage retrieval model.}
\begin{table}[t]
\centering
    \resizebox{\columnwidth}{!}
    {
\begin{tabular}{@{}l|ccccc@{}}
\toprule
 & \begin{tabular}[c]{@{}c@{}}COCO-\\ DR\\ Large \\ (Init.)\end{tabular} & MonoT5 & RankT5 & \begin{tabular}[c]{@{}c@{}}ListT5\\ (r=1)\end{tabular} & \begin{tabular}[c]{@{}c@{}}ListT5\\ (r=2)\end{tabular} \\ \midrule
\begin{tabular}[c]{@{}l@{}}MSMARCO\\ Top-1000\\ (in-domain)\end{tabular} & 41.9 & 43.1 & 46.2 & 46.1 & \textbf{46.3} \\ \midrule
TREC-COVID & 80.8 & \textbf{83.5} & \textbf{83.5} & 83.2 & \textbf{83.5} \\
NFCorpus & 35.5 & 35.6 & 35.5 & \textbf{36.2} & \textbf{36.2} \\
NQ & 54.3 & 57.9 & 59.6 & 59.7 & \textbf{60.0} \\
HotpotQA & 63.3 & 68.7 & \textbf{71.1} & 70.3 & 70.9 \\
FiQA-2018 & 32.3 & 41.2 & 41.3 & \textbf{41.7} & \textbf{41.7} \\
Arguana & 46.9 & 33.0 & 34.8 & 49.0 & \textbf{49.3} \\
Touche-2020 & 21.6 & 25.7 & \textbf{35.7} & 29.1 & 29.6 \\
CQADupStack & 37.3 & 40.5 & 38.7 & 40.7 & \textbf{40.9} \\
Quora & \textbf{87.3} & 84.0 & 83.0 & 86.2 & 86.3 \\
DBPedia & 40.7 & 44.4 & \textbf{46.1} & 45.6 & 45.4 \\
SCIDOCS & 17.3 & 17.5 & 17.5 & 17.7 & \textbf{18.3} \\
FEVER & 74.9 & 78.9 & 79.7 & 79.8 & \textbf{81.4} \\
Climate-FEVER & 23.1 & 24.2 & 22.9 & 23.9 & \textbf{24.9} \\
SciFact & 71.9 & 73.5 & 73.6 & \textbf{74.4} & 74.3 \\ \midrule
Avg. BEIR & 49.1 & 50.6 & 51.6 & 52.7 & \textbf{53.1} \\ \bottomrule
\end{tabular}
}
\caption{NDCG@10 results for reranking on top of COCO-DR large Top-100 first-stage retrieval results (Sec.~\ref{appendix/result_cocodr}).}
\label{table/appendix_cocodr}
\end{table}

\label{appendix/result_cocodr}
We also conduct reranking using the Top-100 results of COCO-DR-large~\cite{cocodr}, and compare reranked results with MonoT5~\cite{monot5}, RankT5~\cite{rankt5}, and \textsc{ListT5}. For this experiment, we could only utilize the evaluation subset that is uploaded (fully open) in the BEIR repository.\footnote{Thus, results for BioASQ, Signal-1M, TREC-NEWS, Robust04 are not computed.} For simplicity, the input token length is fixed to 512 for first-stage retrieval with COCO-DR.
From the results at Tab~\ref{table/appendix_cocodr}, \textsc{ListT5} achieves an average of +1.5 point gain for reranking on COCO-DR Top-100, additionally showing its applicability and effectiveness on both lexical-based (BM25) and neural-based first-stage retrieval modules.

\section{MRR scores}
\label{appendix:mrr_scores}
Table \ref{table/appendix_mrr} reports the corresponding MRR scores for BEIR evaluation on the main table.
\begin{table*}[t!]
\small
\resizebox{\textwidth}{!}{

\begin{tabular}{@{}ll|ccccc|ccccc@{}}
\toprule
\multicolumn{2}{r|}{First-stage retrieval by:} &
  \multicolumn{5}{c|}{BM25 Top100} &
  \multicolumn{5}{c}{COCO-DR Large Top100} \\ \midrule
 &
   &
  \multicolumn{3}{c}{Baselines} &
  \multicolumn{2}{c|}{Ours} &
  \multicolumn{3}{c}{Baselines} &
  \multicolumn{2}{c}{Ours} \\ \cmidrule(lr){3-5} \cmidrule(lr){6-7} \cmidrule(lr){8-10} \cmidrule(l){11-12} 
\multicolumn{1}{c}{Task (Domain)} &
  \multicolumn{1}{c|}{Dataset} &
  \begin{tabular}[c]{@{}c@{}}BM25\\ (Initial)\end{tabular} &
  MonoT5 &
  RankT5 &
  \begin{tabular}[c]{@{}c@{}}ListT5\\ (r = 1)\end{tabular} &
  \begin{tabular}[c]{@{}c@{}}ListT5\\ (r = 2)\end{tabular} &
  \begin{tabular}[c]{@{}c@{}}COCO-DR\\ (Initial)\end{tabular} &
  MonoT5 &
  RankT5 &
  \begin{tabular}[c]{@{}c@{}}ListT5\\ (r = 1)\end{tabular} &
  \begin{tabular}[c]{@{}c@{}}ListT5\\ (r = 2)\end{tabular} \\ \midrule
\multicolumn{1}{l|}{\multirow{4}{*}{\begin{tabular}[c]{@{}l@{}}Passage\\ Retrieval\end{tabular}}} &
  MS MARCO &
  18.0 &
  34.9 &
  \textbf{35.8} &
  35.6 &
  35.8 &
  36.0 &
  38.2 &
  \textbf{40.5} &
  40.3 &
  40.2 \\
\multicolumn{1}{l|}{} &
  \multicolumn{1}{c|}{+ top1000} &
   18.0 &
  37.3 &
  38.6 &
  38.5 &
  \textbf{38.8} &
  36.0 &
  37.2 &
  \textbf{40.3} &
  40.1 &
  40.2 \\
\multicolumn{1}{l|}{} &
  TREC-DL19 &
  82.3 &
  98.3 &
  100.0 &
  98.5 &
  98.8 &
   &
   &
   &
   &
   \\
\multicolumn{1}{l|}{} &
  TREC-DL20 &
  82.4 &
  93.8 &
  95.4 &
  \textbf{95.8} &
  95.6 &
   &
   &
   &
   &
   \\ \midrule
\multicolumn{2}{c|}{\textbf{In-domain average}} &
  61.0 &
  66.1 &
  \textbf{67.4} &
  67.1 &
  67.2 &
  36.0 &
  37.7 &
  \textbf{40.4} &
  40.2 &
  40.2 \\ \midrule
\multicolumn{1}{l|}{\multirow{3}{*}{\begin{tabular}[c]{@{}l@{}}Bio-Medical\\ Information\\ Retrieval\end{tabular}}} &
  TREC-COVID &
  85.3 &
  \textbf{95.4} &
  93.0 &
  93.8 &
  95.0 &
  96.7 &
  96.7 &
  93.8 &
  \textbf{97.0} &
  95.5 \\
\multicolumn{1}{l|}{} &
  NFCorpus &
  52.4 &
  58.0 &
  56.3 &
  57.5 &
  \textbf{58.5} &
  56.0 &
  55.9 &
  57.1 &
  58.1 &
  \textbf{58.4} \\
\multicolumn{1}{l|}{} &
  BioASQ &
  59.4 &
  64.9 &
  \textbf{67.8} &
  67.6 &
  67.2 &
   &
   &
   &
   &
   \\ \midrule
\multicolumn{1}{l|}{\multirow{3}{*}{\begin{tabular}[c]{@{}l@{}}Question\\ Answering\\ (QA)\end{tabular}}} &
  NQ &
  26.3 &
  48.2 &
  \textbf{49.8} &
  48.2 &
  49.5 &
  49.2 &
  52.6 &
  54.7 &
  54.6 &
  \textbf{55.1} \\
\multicolumn{1}{l|}{} &
  HotpotQA &
  80.3 &
  85.9 &
  88.3 &
  88.2 &
  \textbf{88.7} &
  81.9 &
  84.7 &
  87.9 &
  87.6 &
  \textbf{88.5} \\
\multicolumn{1}{l|}{} &
  FiQA-2018 &
  29.6 &
  47.7 &
  48.5 &
  \textbf{48.7} &
  48.5 &
  39.1 &
  49.2 &
  \textbf{49.8} &
  49.6 &
  \textbf{49.8} \\ \midrule
\multicolumn{1}{l|}{Tweet Retrieval} &
  Signal-1M (RT) &
  \textbf{57.4} &
  52.6 &
  55.0 &
  55.6 &
  55.4 &
   &
   &
   &
   &
   \\ \midrule
\multicolumn{1}{l|}{\multirow{2}{*}{News Retrieval}} &
  TREC-NEWS &
  73.9 &
  74.8 &
  75.8 &
  77.9 &
  \textbf{79.2} &
   &
   &
   &
   &
   \\
\multicolumn{1}{l|}{} &
  Robust04 &
  67.3 &
  81.5 &
  \textbf{82.0} &
  79.6 &
  80.0 &
   &
   &
   &
   &
   \\ \midrule
\multicolumn{1}{l|}{\multirow{2}{*}{\begin{tabular}[c]{@{}l@{}}Argument\\ Retrieval\end{tabular}}} &
  Arguana &
  32.8 &
  27.1 &
  27.7 &
  \textbf{40.7} &
  39.6 &
  39.2 &
  26.8 &
  28.6 &
  40.5 &
  \textbf{40.8} \\
\multicolumn{1}{l|}{} &
  Touche-2020 &
  \textbf{74.4} &
  51.3 &
  62.9 &
  58.5 &
  54.9 &
  45.4 &
  47.0 &
  \textbf{61.9} &
  51.1 &
  51.2 \\ \midrule
\multicolumn{1}{l|}{\multirow{2}{*}{\begin{tabular}[c]{@{}l@{}}Duplicate Q.\\ Retrieval\end{tabular}}} &
  CQADupStack &
  29.6 &
  38.5 &
  36.8 &
  38.3 &
  \textbf{38.7} &
  36.9 &
  40.2 &
  38.4 &
  40.7 &
  \textbf{40.8} \\
\multicolumn{1}{l|}{} &
  Quora &
  77.9 &
  83.1 &
  81.2 &
  85.1 &
  \textbf{85.2} &
  \textbf{86.4} &
  82.0 &
  80.5 &
  84.9 &
  84.7 \\ \midrule
\multicolumn{1}{l|}{Entity Retrieval} &
  DBPedia &
  58.2 &
  73.4 &
  74.4 &
  75.7 &
  \textbf{76.3} &
  74.2 &
  73.4 &
  77.2 &
  76.8 &
  \textbf{77.8} \\ \midrule
\multicolumn{1}{l|}{Citation Pred.} &
  SCIDOCS &
   26.0 &
  29.5 &
  29.9 &
  30.9 &
  \textbf{31.2} &
  30.9 &
  30.2 &
  30.6 &
  31.1 &
  \textbf{32.0} \\ \midrule
\multicolumn{1}{l|}{\multirow{3}{*}{Fact Checking}} &
  FEVER &
   62.4 &
  78.4 &
  77.6 &
  77.6 &
  \textbf{80.2} &
  74.2 &
  79.0 &
  80.1 &
  80.3 &
  \textbf{82.1} \\
\multicolumn{1}{l|}{} &
  Climate-FEVER &
  22.0 &
  31.9 &
  29.1 &
  31.3 &
  \textbf{33.7} &
  32.9 &
  33.0 &
  31.3 &
  33.0 &
  \textbf{34.8} \\
\multicolumn{1}{l|}{} &
  SciFact &
  64.6 &
  70.3 &
  70.6 &
  71.8 &
  \textbf{71.9} &
  68.4 &
  70.9 &
  70.4 &
  71.8 &
  \textbf{71.8} \\ \midrule
\multicolumn{2}{c|}{\textbf{BeIR average}} &
  54.4 &
  60.7 &
  61.5 &
  62.6 &
  \textbf{63.0} &
  58.0 &
  58.7 &
  60.2 &
  61.2 &
  \textbf{61.7} \\ \bottomrule
\end{tabular}

}
\caption{Reporting Mean Reciprocal Rank (MRR)@10 scores for initial ranking of BM25 top-100 and COCO-DR large top-100. (Sec.~\ref{appendix:mrr_scores})}
\label{table/appendix_mrr}
\end{table*}

\section{Hyperparameter Optimization Methodology}
\label{appendix/hyperparameter}
We conduct a structured grid search strategy on a selected subset of the BEIR dataset. The selection of hyperparameters for investigation was informed by prior research and best practices within the domain, specifically referencing the work by \cite{rankt5}, which demonstrated effective use of a learning rate of $1 \times 10^{-4}$. To explore the potential for further optimization, we extended the search to include lower learning rates down to $1 \times 10^{-5}$, considering the sensitivity of large models to learning rate adjustments.

We employed a linear learning rate scheduler with a 10\% warmup over 10 epochs, a strategy for stabilizing training in the initial phases. The grid search was designed with a step size range from 10,000 to 30,000, increasing in increments of 10,000 steps.

For our 3 billion parameter models (3B models), due to their substantial computational requirements, we conducted a more granular evaluation. We assessed performance at every 1,000 steps between 1,000 and 10,000 steps.

This structured approach ensured that our hyperparameter selection was not arbitrary but based on a reasoned strategy aiming to balance computational efficiency with empirical performance. The specific choices of "250k steps" at a "1e-4" learning rate and "3k steps" at a "1e-5" learning rate were outcomes of this optimization process, identified as configurations that provided the best performance metrics within the constraints of our experimental setup.

\section{Details on Evaluation Dataset}
\label{appendix/evaluation}
\subsection{Baseline Models.}
We download and use the officially released checkpoints for the baseline model with the huggingface identifier of \texttt{castorini/monot5-base-msmarco-10k, monot5-3b-msmarco-10k} for MonoT5~\cite{monot5}, \texttt{castorini/duot5-base-msmarco} for DuoT5~\cite{duot5}, \texttt{castorini/rank\_vicuna\_7b\_v1} for RankVicuna~\cite{rankvicuna}, and \texttt{castorini/rank\_zephyr\_7b\_v1\_full} for RankZephyr~\cite{rankzephyr}.
For RankGPT\cite{rankgpt} using GPT-3.5 or GPT-4, we evaluate them using the OpenAI~\cite{openai} API. 
For RankT5, we download the released T5X checkpoint from the official repository\footnote{\url{https://github.com/google-research/google-research/tree/master/rankt5}} and convert it into a PyTorch checkpoint using the HuggingFace's conversion script\footnote{convert\_t5x\_checkpoint\_to\_pytorch.py}. We evaluate DuoT5 on the same setup to the paper's final experiments - we rerank 100 passages using MonoT5, select top 50 passages to be run on DuoT5, and aggregate relevancy scores of individual documents by the SYM-SUM method.

We run the official RankLLM repository\footnote{\url{https://github.com/castorini/rank_llm}} to evaluate RankVicuna and RankZephyr on BEIR.

\subsection{Links, Maximum Input Length}
Since the average length of query and passage differs greatly for each BEIR subset, we assign the appropriate sequence length for each BEIR dataset and evaluate all models in the same setup. We download the dumped index of the top 100 retrieved passages by BM25 from the official Pyserini repository\footnote{\url{https://github.com/castorini/pyserini}}, and download the inital query, corpus, and qrels from the BEIR repository.\footnote{\url{https://public.ukp.informatik.tu-darmstadt.de/thakur/BEIR/datasets/}}

We referenced the average query + passage length from the official BEIR paper. We selected the smallest maximum length from [256, 512, and 1024], but is still bigger than the (sum of average query + passage length) multiplied by two(for better coverage). One exception is the trec-news, which added up to be bigger than 1024, but we just capped it to be 1024. The \textit{exact} alias with the maximum input length is listed below;  

`msmarco': 256, `dl19': 256, `dl20': 256, `trec-covid': 512, `nfcorpus': 512, `bioasq': 512, `nq': 256, `hotpotqa': 256, `fiqa': 512, `signal': 256, `news': 1024, `robust04': 1024, `arguana': 1024, `touche': 1024, `cqadupstack': 512, `quora': 256, `dbpedia-entity': 256, `scidocs': 512, `fever': 256, `climate-fever': 256, `msmarco\_top1000': 256, `scifact': 512

\subsection{Other Replication Details}
\textbf{Output validation module} Upon experiments, we do not use constrained decoding or validation module to ensure the generation of valid permutations of 5. Each digital number is represented as simple integers (The decoded output looks something like "1 2 5 4 3" or "2 3 1 5 4"), with an exception catching module such that if the parsing fail, we go back to the original ordering. Even without any validation module, since \textsc{ListT5} is fine-tuned to output valid indexes for 20000 steps, if the input follows the correct format, the model almost always outputs valid index. Nevertheless, we think that adding a constrained decoding module would be beneficial.

\textbf{Evaluation} We run the same evaluation setup with the BEIR library to do evaluation. For example, we append title with a space along with the text(passage) part and treat as full passage. For evaluating reranked results for both BM25 Top100 and COCO-DR Top100 candidate passages, we remove duplicates before evaluation\footnote{For example, quora had 1 (out of 10000) duplicates, so the NDCG@10 for BM25 may differ from the BEIR paper.} For CQADupStack, we follow the same procedure and aggregated all sub-domains ranging from android to wordpress to make 13,145 test queries, and report the average performance.

\section{Details about the Experiment Measuring Positional Bias.}
\label{ablation/permutation_consistency}
\subsection{Evaluation process.} We briefly explain here about the evaluation process for the FiQA dataset. We defined the positional bias inspired from Liu et al., [9], where they measure the answer accuracy with respect to the index change of the relevant passage. 

1. We break down the original dataset into pairs of one query and one positive passage associated with the query. For each query-positive pair, we randomly sample 4 negative passages from the BM25 top100 results and additionally pair them to make one sample. Preprocessing them makes 818 distinct samples.

2. For every sample, we make 5 variants of the input text, so that the first one assigns index 1 to the positive passage, and the last one assigns index 5 to the positive passage. Note that only the index of the positive passage is changed, and we keep the order of other passages (negatives) as the same. 

3. we forward the inputs to each model, and collect the output that corresponds to the most relevant index.
To answer (1), we measure the accuracy with respect to the index of positive passage. To answer (2), we analyze the ratio of samples where the model points to the same passage regardless of positive index change.

We discard queries that doesn't have positive indexes in the bm25 top100 dataset, or those that don't have 4 distinct negative contexts.
For reproducibility, all randomly sampled datasets used for experiments are conducted with fixed seed. (seed=0) Standard deviation are calculated using the numpy.std function.

\subsection{Prompt example.}

To enforce a scenario where the model takes lengthy inputs at once, we apply the instruction format of permutation generation (text), and use the same prompt described in the RankVicuna \cite{rankvicuna} paper.
The prompts we use to give inputs to GPT-3.5-turbo-0301 are the following:

\begin{small}
\begin{verbatim}
I will provide you with 5 passages, 
each indicated by numerical identifier [].
Rank the passages based on their 
relevance to the search query: {query}.

[1] {passage_1}
[2] {passage_2}
[3] {passage_3}
[4] {passage_4}
[5] {passage_5}

Search Query: {query}

Rank the 5 passages above based on their 
relevance to the search query.
All the passages should be included and 
listed using identifiers, in descending 
order of relevance. The output format 
should be [] > [], e.g., [4] > [2].
Only respond with the ranking 
results, do not say any word or explain.             
\end{verbatim}
\end{small}

\noindent Given the above input, we collect the output of the LLM. Following the setup from \citet{rankgpt}, we use the ChatCompletion api by the openai library, with temperature of 0. Using the openai api, running the whole process cost about \$35. To run ths GPT4 experiment on FiQA, it additionally cost 195.8\$ (total of 6523910 input / 1298 output tokens).
After the run, we found that 3 out of 818 outputs had malformed outputs, in a format such as: "[No relevant passage found] > ... [1]", rather than in a correct format such as: "[3] > [1] > [5] > [2] > [4]".
We discard those 3 malformatted instances with incorrect output formats into consideration and compared only with the rest. (That is, we also discard those instances for the FiD-T5 variant)
To give an exactly same scenario to both models, we truncate each passages up to 512 tokens (based on the T5 tokenizer), since that is exactly what our fid model sees. After truncation, we also validated that all of the input sizes are below the range of 4096 (as the maximum input window size of gpt-3.5-turbo-0301 is 4096)
\subsection{Detailed analysis about pairwise methods}
We find that pairwise models also exhibit positional bias, having a tendency to label the passage that comes at the front as positive.
Theoretically, we can remove the positional sensitivity by evaluating all possible permutations of input passages \cite{pairwise_murphy, pairwise_yarotsky}. DuoT5 and other pairwise ranking methods~\cite{llms-effective-rankers} already include swapping orders and aggregate scores from both orderings. To measure the positional bias of pairwise models, in this experiment, we removed the averaging from swapping orders in DuoT5. However, the number of forward passes needed to mitigate positional bias problems in this way grows in a factorial scale, making it impossible and impractical to be applied to listwise methods. (We need 5! = 120 number of forward passes in order to remove the positional bias of 5 instances) In contrast, we effectively mitigate the positional bias problem, without any additional forward passes, with the Fusion-in-Decoder architecture.
\begin{table*}[t!]
\small
\resizebox{\textwidth}{!}{

\begin{tabular}{@{}lcc|l@{}}
\toprule
\multicolumn{1}{l|}{} & \multicolumn{1}{c|}{GT Rank / GT Score} & 1st Rank / 1st Score & \multicolumn{1}{c}{Passage} \\ \midrule
\multicolumn{1}{l|}{\textsc{ListT5}} & \multicolumn{1}{c|}{1} & 1 & \begin{tabular}[c]{@{}l@{}}Most often, prosecutions that occur are not just with only the \textbf{losing side} being \red{prosecuted for} \\ \red{their crimes}. The Nuremburg trials prosecuted Nazi’s for offences they committed, but none of the\\  Allied forces were ever brought for trial(...)\end{tabular} \\ \midrule
\multicolumn{1}{l|}{MonoT5} & \multicolumn{1}{c|}{78/ -0.32} & 1 / -0.13 & \begin{tabular}[c]{@{}l@{}}As the \blue{ICC} intentionally limits its \red{prosecutions} to group leaders, many of those who actually commit \\ atrocities need have no fear of prosecution  By prosecuting only those leaders deemed ‘most responsible’\\  for the crimes in question, the \blue{ICC} is effectively allowing lower-ranked perpetrators to commit crimes \\ with impunity. (...)\end{tabular} \\ \midrule
\multicolumn{1}{l|}{RankT5} & \multicolumn{1}{c|}{25/ -5.86} & 1 / -4.58 & \begin{tabular}[c]{@{}l@{}}\blue{Accountability}  It is a fundamental principle of morality that individuals should be held responsible\\  for their crimes – that is the reason why we, as societies, have criminal law.  \red{Prosecuting} people \\ – holding them responsible for their crimes – is a \blue{moral imperative}. We all wish to live in a \\ society where everyone is equally accountable when they commit crime as one in which not everyone\\  is held to account is fundamentally unjust (...)\end{tabular} \\ \midrule
\multicolumn{3}{c|}{Query} & \begin{tabular}[c]{@{}l@{}}(...) \red{Prosecuting offenders} is the \textbf{only way to get a just outcome} when there have been horrific \\ crimes committed. At a most principled level, those who commit a crime ought to be held accountable\\  for their actions even if they are powerful or it damages the chances of peace because the powerful\\  must be shown not to be above the law (...)\end{tabular} \\ \midrule
\multicolumn{3}{c|}{Relevant Passage} & \begin{tabular}[c]{@{}l@{}}Most often, prosecutions that occur are not just with only the \textbf{losing side} being \red{prosecuted for} \\ \red{their crimes}. The Nuremburg trials prosecuted Nazi’s for offences they committed, but none of the\\  Allied forces were ever brought for trial(...)\end{tabular} \\ \bottomrule
\end{tabular}

}
\caption{Example of ListT5, MonoT5, RankT5 retrieval result on Arguana dataset. Relevant topics are highlighted with \red{red}, non-relevant topics in \blue{blue}, and opinions are \textbf{bolded}. (Sec.~\ref{appendix/qualitative})}
\label{table/appendix_Qualitative_analysis}
\end{table*}

\subsection{Experiments on the consistency of LLMs.}
\begin{table}
    \resizebox{\columnwidth}{!}
    {
\begin{tabular}{@{}lccccc@{}}
\multicolumn{1}{c|}{} &
  \multicolumn{4}{c|}{GPT-3.5-turbo-1106} &
  \multirow{2}{*}{\begin{tabular}[c]{@{}c@{}}ListT5\\ -base\end{tabular}} \\
\multicolumn{1}{c|}{}      & Trial 1 & Trial 2 & Trial 3 & \multicolumn{1}{c|}{Avg.}            &                 \\ \midrule
\multicolumn{6}{l}{(1) Accuracy when the gold passage is at index \#:}                                            \\ \midrule
\multicolumn{1}{l|}{1}     & 81.6   & 79.6   & 81.6   & \multicolumn{1}{c|}{\textbf{81.0 }} & \textbf{93.9 } \\
\multicolumn{1}{l|}{2}     & 63.3   & 63.3   & 61.2   & \multicolumn{1}{c|}{\textbf{62.6 }} & \textbf{87.8 } \\
\multicolumn{1}{l|}{3}     & 75.5   & 75.5   & 75.5   & \multicolumn{1}{c|}{\textbf{75.5 }} & \textbf{83.7 } \\
\multicolumn{1}{l|}{4}     & 67.3   & 63.3   & 67.3   & \multicolumn{1}{c|}{\textbf{66.0 }} & \textbf{85.7 } \\
\multicolumn{1}{l|}{5}     & 61.2   & 63.3   & 65.3   & \multicolumn{1}{c|}{\textbf{63.3 }} & \textbf{81.6 } \\
\multicolumn{1}{l|}{std}   & 7.68   & 7.1    & 7.4    & \multicolumn{1}{c|}{\textbf{7.4 }}  & \textbf{4.2 }  \\ \midrule
\multicolumn{6}{l}{(2) Agreement ratio (\%) within index change of positive}                                      \\ \midrule
\multicolumn{1}{l|}{\begin{tabular}[c]{@{}l@{}}points to \\ same passage\end{tabular}} &
  55.1  &
  55.1  &
  55.1  &
  \multicolumn{1}{c|}{\textbf{55.1 }} &
  \textbf{83.7 } \\ \midrule
\multicolumn{1}{l|}{other} & 44.9   & 44.9   & 44.9   & \multicolumn{1}{c|}{44.9 }          & 16.3          
\end{tabular}

}
\caption{Measuring the LLM consistency on TREC-COVID. (Sec.~\ref{appendix:llm_consistency})}
\label{table/llm_consistency}

\end{table}


\label{appendix:llm_consistency}
We ran the positional bias experiment (on GPT3.5) on TREC-COVID for 3 times to investigate the following: given the same input, does LLMs generate the same output multiple times? \footnote{We were only able to run the inconsistency experiment on GPT3.5 due to the high inference cost of GPT-4.} The results from api calls at Tab.~\ref{table/llm_consistency} were not always exactly the same, but the difference was negligible, not changing the main claims. For the case of \textsc{ListT5}, it is deterministic and thus fully replicable. On running the same experiments on FiQA twice, we validated that the output files for \textsc{ListT5} are exactly the same.

\section{Qualitative Analysis}
\label{appendix/qualitative}
Tab.~\ref{table/appendix_Qualitative_analysis} reports 
a sample output selected from the Arguana Dataset, which seeks the counter-argument most opposed to the query argument .\footnote{The query is always included in the corpus, so for qualitative analysis, we excluded the passage that is exactly the same with the original query.} Considering the NDCG@10 reranking performance for MonoT5-base, RankT5-base, and \textsc{ListT5}-base is 34.4, 35.5, and 48.9, respectively, \textsc{ListT5} excels on Arguana than other pointwise baseline models with a large margin. From inspection, we conclude that the listwise reranking nature of \textsc{ListT5} helps to discern and select which features are important in determining relevancy in this dataset, with respect to other passages. 

We hypothesize that listwise comparison makes the model to favor counterarguments - (Different point of view, but the discussion topic is the same) over similar topics. For example at Tab.~\ref{table/appendix_Qualitative_analysis}, RankT5 places passages that talks about accountability and criminal law in a moral aspect as the most relevant passage, which is quite similar to the original query (since it includes words such as `Prosecuting') but not exactly the same about prosecuting offenders. Also, the passage that is scored highest from MonoT5 talks about ICC's prosecution policy, but the original query is not related to ICC. In contrast, since perhaps \textsc{ListT5} is able to see different passages, it correctly determines the GT passage as top-1, giving more weight on talking about the same topic (even though the aspect can be different). 

\begin{table*}[t]

\resizebox{\textwidth}{!}
{

\begin{tabular}{@{}l|l|l@{}}
\toprule
Sorting method & \# required forward passes to rerank top1 & \# required forward passes to rerank top10 \\ \midrule
sliding window, stride=1 & 1 + {$\lceil$(100-5)/1$\rceil$} = 96           & 96 $\times$ {$\lceil$10/4$\rceil$} = 288 -> 280  \\
sliding window, stride=2 & 1 + $\lceil$(100-5)/2$\rceil$ = 49           & 49 $\times$ {$\lceil$10/3$\rceil$} = 196 -> 191 \\
sliding window, stride=3 & 1 + {$\lceil$(100-5)/3$\rceil$} = 33           & 33 $\times$ {$\lceil$10/2$\rceil$} = 165 -> 162    \\
sliding window, stride=4 & 1 + {$\lceil$(100-5)/4$\rceil$} = \textbf{25}  & 25 $\times$ {$\lceil$10/1$\rceil$} = 250 -> 248   \\ \midrule
tournament sort, r=1     & (100/5) + (20/5) + 1 = \textbf{25} & 25 + 9 x (1+1+1) = \textbf{52}   \\
tournament sort, r=2     & (100/5) + (40/5) + 2 + 1 = 31      & 31 + 9 $\times$ (1+1+1+1) = 67          \\ \bottomrule
\end{tabular}

}
\caption{Number of forward passes to rerank top-k candidates from 100 candidate passages per one query, where window size $w$=5. In the case of reranking top-10 passages, tournament sort requires much more fewer number of forward passes. (Sec.~\ref{appendix:required_forward})}
\label{table/appendix_efficiency_numforwards}

\end{table*}

\begin{table}

    \resizebox{\columnwidth}{!}
    {

\begin{tabular}{@{}lcll|cc@{}}
\toprule
\multirow{2}{*}{Idx} & \multirow{2}{*}{Base Model} & \multirow{2}{*}{\begin{tabular}[c]{@{}l@{}}Sorting\\ method\end{tabular}} & \multirow{2}{*}{Name} & \multicolumn{2}{c}{FLOPs to rerank:} \\ \cmidrule(l){5-6} 
 &  &  &  & \multicolumn{1}{c|}{Top-1} & Top-10 \\ \midrule
0 & T5-base & pointwise & MonoT5 & \multicolumn{1}{c|}{1x} & 1x \\
1 & T5-base & tournament & ListT5(r=1) & \multicolumn{1}{c|}{1.3x} & 2.6x \\
2 & T5-base & tournament & ListT5(r=2) & \multicolumn{1}{c|}{1.8x} & 4.7x \\
3 & T5-base & sliding w.(s=2) & T5(FiD) & \multicolumn{1}{c|}{2.5x} & 9.8x \\
4 & T5-base & sliding w.(s=3) & T5(FiD) & \multicolumn{1}{c|}{1.7x} & 12.3x \\ \midrule
5 & T5-3b & tournament & ListT5(r=1) & \multicolumn{1}{c|}{17.6x} & 36.3x \\
6 & T5-3b & tournament & ListT5(r=2) & \multicolumn{1}{c|}{24.6x} & 66.0x \\
7 & T5-3b & sliding w.(s=2) & T5(FiD) & \multicolumn{1}{c|}{38.5x} & 154x \\
8 & T5-3b & sliding w.(s=2) & T5(no FiD) & \multicolumn{1}{c|}{53.8x} & 215.1x \\
9 & T5-3b & sliding w.(s=3) & T5(FiD) & \multicolumn{1}{c|}{25.6x} & 128x \\
10 & T5-3b & sliding w.(s=3) & T5(no FiD) & \multicolumn{1}{c|}{35.1x} & 175.6x \\ \bottomrule
\end{tabular}

}
\caption{FLOPs (In a multiple of FLOPs of MonoT5-base) on the choice of architecture and method, on TREC-DL19. For the sliding window approach, we would need a total of 4 multiple passes for stride = 3 and 5 passes for stride = 2 (Explained at Tab.~\ref{table/appendix_efficiency_numforwards}) to rerank Top-10 candidates. (Sec.~\ref{appendix:flops_comp})}
\label{table/appendix_efficiency_flops}

\end{table}


\begin{table}[t!]
\centering
    \resizebox{\columnwidth}{!}
    {
\begin{tabular}{@{}lcccc|cccc@{}}
\toprule
\multicolumn{1}{l|}{} & \multicolumn{4}{c|}{Rerank Top-10 (NDCG@10)} & \multicolumn{4}{c}{Rerank Top-1 (NDCG@1)} \\ \midrule
\multicolumn{1}{l|}{Sorting Method} & \multicolumn{2}{c}{T.S.} & \multicolumn{2}{c|}{S.W} & \multicolumn{2}{c}{T.S.} & \multicolumn{2}{c}{S.W.} \\ \midrule
\multicolumn{1}{l|}{Hyperparam.} & r=1 & r=2 & \begin{tabular}[c]{@{}c@{}}s=2\\ (iter=5)\end{tabular} & \begin{tabular}[c]{@{}c@{}}s=3\\ (iter=4)\end{tabular} & r=1 & r=2 & \begin{tabular}[c]{@{}c@{}}s=2\\ (iter=1)\end{tabular} & \begin{tabular}[c]{@{}c@{}}s=3\\ (iter=1)\end{tabular} \\ \midrule
\multicolumn{1}{l|}{FLOPS(DL19)} & \textbf{1x} & 1.8x & 3.7x & 3.1x & \textbf{1x} & 1.4x & 1.96x & 1.32x \\ \midrule
\multicolumn{1}{l|}{DL19} & 71.2 & 71.8 & 71.5 & 71.8 & 81.0 & 79.1 & 81.0 & 78.7 \\
\multicolumn{1}{l|}{DL20} & 67.3 & 68.1 & 67.3 & 67.7 & 77.8 & 77.8 & 79.0 & 79.6 \\ \midrule
In-domain avg. & 69.3 & \textbf{70.0} & 69.4 & 69.8 & 79.4 & 78.5 & \textbf{80.0} & 79.2 \\ \midrule
\multicolumn{1}{l|}{TREC-COVID} & 76.7 & 78.3 & 78.9 & 77.5 & 88.0 & 91.0 & 88.0 & 86.0 \\
\multicolumn{1}{l|}{NFCorpus} & 35.5 & 35.6 & 35.3 & 35.5 & 47.8 & 49.2 & 48.6 & 48.6 \\
\multicolumn{1}{l|}{BioASQ} & 57.2 & 56.4 & 54.5 & 54.9 & 59.2 & 58.4 & 55.8 & 57.2 \\
\multicolumn{1}{l|}{NQ} & 52.0 & 53.1 & 52.7 & 52.8 & 36.0 & 37.6 & 36.4 & 36.6 \\
\multicolumn{1}{l|}{HotpotQA} & 72.1 & 72.6 & 71.2 & 71.6 & 83.3 & 84.1 & 83.1 & 83.1 \\
\multicolumn{1}{l|}{FiQA-2018} & 39.5 & 39.6 & 39.7 & 39.8 & 41.2 & 40.7 & 41.4 & 41.5 \\
\multicolumn{1}{l|}{Signal-1M (RT)} & 33.3 & 33.5 & 32.4 & 33.2 & 43.3 & 41.8 & 42.3 & 41.8 \\
\multicolumn{1}{l|}{TREC-NEWS} & 47.9 & 48.5 & 49.8 & 50.0 & 53.2 & 54.1 & 52.3 & 52.9 \\
\multicolumn{1}{l|}{Robust04} & 52.0 & 52.1 & 51.3 & 51.7 & 65.1 & 66.3 & 67.1 & 65.9 \\
\multicolumn{1}{l|}{Arguana} & 49.7 & 48.9 & 47.7 & 47.8 & 25.8 & 23.9 & 23.3 & 22.7 \\
\multicolumn{1}{l|}{Touche-2020} & 34.2 & 33.4 & 32.7 & 33.1 & 34.7 & 31.6 & 36.7 & 36.7 \\
\multicolumn{1}{l|}{CQADupStack} & 38.4 & 38.8 & 38.9 & 38.8 & 31.6 & 31.9 & 32.1 & 32.0 \\
\multicolumn{1}{l|}{Quora} & 86.1 & 86.4 & 86.3 & 86.2 & 77.8 & 77.8 & 78.1 & 77.7 \\
\multicolumn{1}{l|}{DBPedia} & 43.9 & 43.7 & 42.6 & 43.2 & 55.5 & 56.5 & 55.1 & 56.6 \\
\multicolumn{1}{l|}{SCIDOCS} & 17.2 & 17.6 & 17.9 & 17.7 & 21.9 & 22.0 & 22.8 & 21.4 \\
\multicolumn{1}{l|}{FEVER} & 77.8 & 79.8 & 79.3 & 79.3 & 69.4 & 72.4 & 70.2 & 70.4 \\
\multicolumn{1}{l|}{Climate-FEVER} & 22.8 & 24.0 & 23.8 & 23.7 & 20.2 & 23.3 & 20.4 & 21.0 \\
\multicolumn{1}{l|}{SciFact} & 74.1 & 74.1 & 73.6 & 73.5 & 65.0 & 65.3 & 65.3 & 65.7 \\ \midrule
\multicolumn{1}{l|}{BEIR avg.} & 50.6 & \textbf{50.9} & 50.5 & 50.6 & 51.1 & \textbf{51.6} & 51.1 & 51.0 \\ \bottomrule
\end{tabular}
}
\caption{Comparison of FLOPs and performance on \textsc{ListT5} with the \textbf{S.W} (Sliding Window) approach and \textbf{T.S} (Tournament Sort) with different hyperparameters. \textsc{ListT5} ($r$=2) performs the best, with lower FLOPs than the sliding window variants on both setup of reranking top-10 and top-1 candidates. (Sec.~\ref{appendix:perf_comp})}
\label{table/appendix_sliding_window}
\end{table}

\section{Measuring efficiency - Sliding window v.s. Tournament Sort.}
\begin{figure*}[!ht]
{
\centering
    \includegraphics[width=0.95\textwidth]{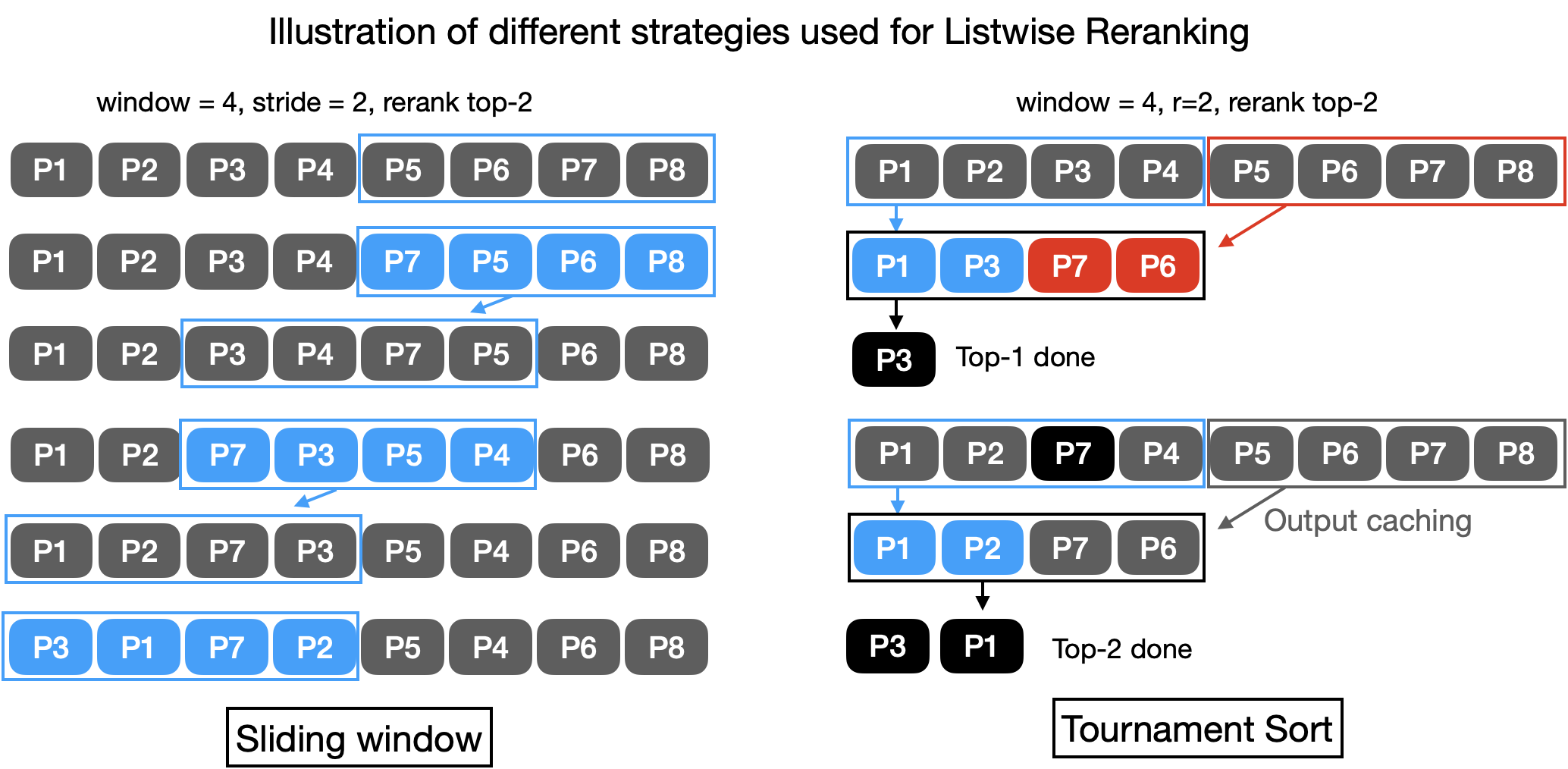}
    \caption{Overview comparison of sliding window v.s. Tournament sort for listwise reranking. (Sec.~\ref{appendix/efficiency})}
    \label{fig:sliding_tournament}
    \vspace{-0.5cm}
}
\end{figure*}

\label{appendix/efficiency}
\label{sec:ablation_efficiency}
We briefly describe the difference of tournament sort and the sliding window approach in Figure.~\ref{fig:sliding_tournament}. We conduct additional experiments to compare the efficiency of tournament sort and the sliding window.

\subsection{Implementation Details.}

We rerank top-10 passages from BM25 top100 candidate passages, for 43 queries of TREC-DL19. The latency and FLOPs for tournament sort variants are computed using our \textsc{ListT5} evaluation code, and the efficiency for the sliding window approach (without the FiD architecture) is computed using the official repository for RankVicuna and RankZephyr.\footnote{\url{https://github.com/castorini/rank_llm}} The FiD variant of the sliding window approach are computed using our \textsc{ListT5} evaluation code. We append the deepspeed FlopsProfiler for FLOPs measurement, which is the same as the ones that was appended for \textsc{ListT5}. For a fair comparison, we also measure the FLOPs of the sliding approach on the T5-3b model\footnote{we replace the model code from castorini/rank\_vicuna\_7b\_v1 to lmsys/fastchat-t5-3b-v1.0 and measure FLOPs.} and the T5-base FiD architecture using the \textsc{ListT5} code. Example commands used are as follows:

\begin{small}
\begin{verbatim}
CUDA_VISIBLE_DEVICES=0 python 
./src/rank_llm/scripts/run_rank_llm.py
--model_path=castorini/lmsys/
fastchat-t5-3b-v1.0
--top_k_candidates=100
--dataset=dl19 --retrieval_method=bm25 
--prompt_mode=rank_GPT --
context_size=4096 --variable_passages 
 --window_size 5 --step_size [2 or 3]
\end{verbatim}
\end{small}

\subsection{Number of Required Forward Passes.}
\label{appendix:required_forward}
We calculate the number of required forward passes needed for variants of tournament sort and the sliding window approach at Table ~\ref{table/appendix_efficiency_numforwards}. Given a window size of 5, tournament sort is much more efficient to rerank top-10 passages, requiring fewer number of forward passes. This is because, unlike tournament sort with output caching, the sliding window approach requires re-evaluation over the entire input sequence, depending on the window size, stride, and the number of top-k passages to rerank. Detailed explanation of how we calculated the numbers from the table is below:

\begin{itemize}
    \item{After one pass of a sliding window of size 5 and stride of 4, we discard 4 passages and only carry (5-4)=1 previous passage to the next step as we move the window. Therefore, we would only be able to correctly order top-1 passages, since the top-2 passage cannot be moved.}
    \item{Therefore, to rank top 10 candidates, we have to iterate through at least $\lceil$10 / (5-4)$\rceil$ = 10 times, which would result in a total of about 250 forward passes.}
    \item{For ranking top-6 with a sliding window of size 5 and stride 4, some forward pass can be saved in the 6th slide because top-5 have already been ordered in the first 5 slides. Therefore, a corrected precise calculation (noted as -> in the table) gives 248.}
    \item{The same applies to methods of stride=2 to 4.}
    \item{In contrast, tournament sort uses output caching, and after the initial computation (25 for r=1 and 31 for r=2), we only need to compute one path from leaf to root, which only costs only one additional forwards for each level of the tournament tree, which is 3 for (r=1) and 4 for (r=2).}
    \item{Therefore, by using tournament sort, we can efficiently reduce the number of forward passes needed to rank top 10 candidates.}
\end{itemize}

\subsection{FLOPs comparision}
\label{appendix:flops_comp}

In this section, we compare in detail on the choice of sorting method and architecture at Tab.~\ref{table/appendix_efficiency_flops}.

\textbf{T5 (FiD) v.s. T5 (no FiD).}
ListT5 uses the FiD architecture to effectively mitigate the positional bias problem and handle long inputs efficiently.
Comparing with idx 7 v.s 8 (38.5 vs 53.8) and 9 vs 10 (25.6 vs 35.1) at Tab.~\ref{table/appendix_efficiency_flops}, we can see that using FiD results in lower number of total FLOPs.

\textbf{Tournament sort v.s. Sliding window on T5 (FiD).}
By comparing idx 5 and 6 with respect to idx 7 and 9, we conclude that the FLOPs to rerank Top-10 candidates are much lower for both ($r$ = 1) and ($r$ = 2) variants of tournament sort (36.3x and 66.0x), compared with the FiD variant of sliding window, for both (stride = 2) and (stride = 3) (154x and 128x). It also holds the same for models built on top of T5-base, by comparing idx (1,2) with respect to (3,4).

\subsection{Performance Comparison.}
\label{appendix:perf_comp}
Comparison of FLOPs and performance on \textsc{ListT5} with the S.W(Sliding Window) approach and T.S(Tournament Sort) with different hyperparameters. \textsc{ListT5} ($r$=2) performs the best, with lower FLOPs than the sliding window variants on both setup of reranking top-10 and top-1 candidates.

\section{Additional Experiments on Tournament Sort.}
\begin{figure}[!t]
{
\centering
    \includegraphics[width=\columnwidth]{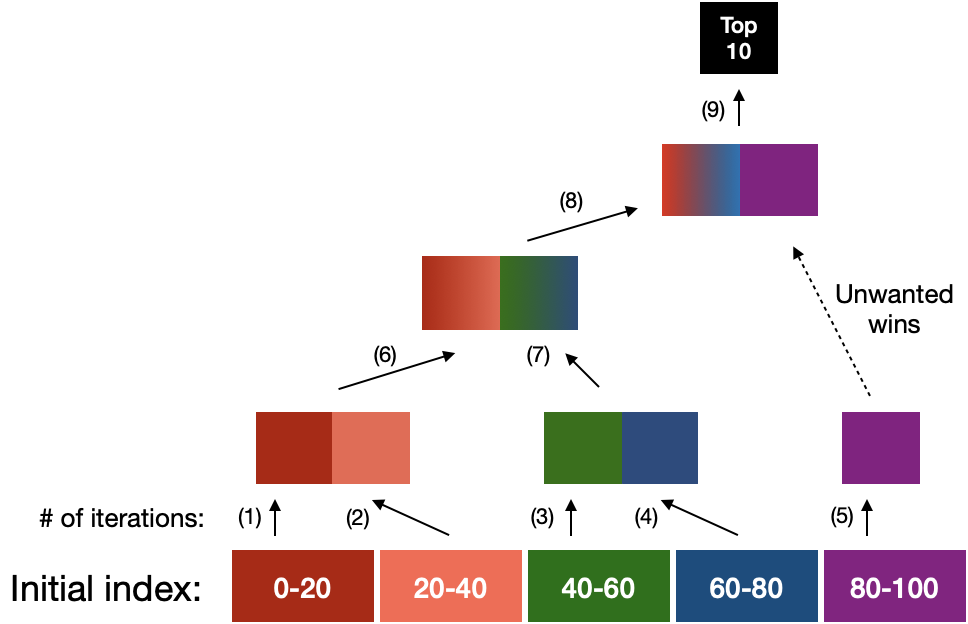}
    \caption{Illustration for the tournament sort to rank top-10 passages among 100 candidates for $r$=10 and $m$=20. It takes 9 forward passes with unwanted wins. (Sec.~\ref{appendix_m_k})}
    \label{fig:fig_tournament_sort_iter}
}
\end{figure}

\begin{table}[t]
\centering
    \resizebox{\columnwidth}{!}
    {
\begin{tabular}{@{}l|lll@{}}
\toprule
 & \multicolumn{3}{c}{\# of stride == value of r} \\ \midrule
 & \multicolumn{1}{c|}{10} & \multicolumn{1}{c|}{5} & \multicolumn{1}{c}{1} \\ \midrule
Sliding w. & \multicolumn{1}{l|}{9 x 1 iter. = \textbf{9}} & \multicolumn{1}{l|}{17 x 2 iter. = 34} & 81 x 10 iter. = 810 \\
Tournament S. & \multicolumn{1}{l|}{9 x 1 iter. = \textbf{9}} & \multicolumn{1}{l|}{7 x 2 iter. = \textbf{14}} & 6 x 10 iter. = 
\textbf{60} \\ \bottomrule
\end{tabular}
}
\caption{Comparison of the number of forward passes needed to rank top-10 passages, when $m$ = 20, and $n$ = 100. The number is written in the format of \{number of forward pass for each iteration\} x \{number of iterations to assure global ranking\}. (Sec.~\ref{appendix_m_k})}
\label{table/appendix_tournament_extension}
\end{table}

\subsection{Cases where m > k}
\label{appendix_m_k}
In the main paper, we have discussed that the sliding window approach can be much more efficient when $m$ << $k$. However, even when $m$ > $k$, e.g., $m$ = 20 and $k$ = 10, the number of forward passes needed to get top-10 rankings for tournament sort can less than or equal to the sliding window variants, with simple modifications. For example, the number of iterations to rank top-10 with $m$=20 and $r$ = 10 with \textsc{ListT5} requires 9 forward passes if we take into account all top-$r$ results from one iteration and take into account unwanted wins. Tournament sort with $r$=10 can correctly rank global top-10 candidates in one iteration, as illustrated at Fig.~\ref{fig:fig_tournament_sort_iter}. This is the same amount of iterations needed with the sliding window approach with window size of 20 and stride of 10. We also compute the number of forward passes with different value of $s$, or $r$ in Table.~\ref{table/appendix_tournament_extension}.

\subsection{Tournament Sort with LLMs.}
\begin{table}[t]
\centering
    \resizebox{\columnwidth}{!}
    {
\begin{tabular}{@{}lccccc@{}}
\toprule
Method & dl19 & dl20 & trec-covid & news & touche \\ \midrule
sliding & 68.4 ± 0.4 & 64.9 ± 1.1 & 72.6 ± 1.4 & 46.5 ± 1.0 & \textbf{38.2 ± 0.5} \\
tournament & 67.4 ± 0.9 & 65.8 ± 0.6 & \textbf{76.4 ± 0.4} & 45.5 ± 1.0 & 33.1 ± 1.7 \\ \bottomrule
\end{tabular}
}
\caption{NDCG@10 on the selected subset of BEIR, on RankGPT-3.5 with different sorting methods. For fair comparison, we used w = 20, s = 10 for the sliding approach, and m = 20, r = 10 for the tournament sort. To compensate for the instability of APIs, all results are run for 3 times. Except for trec-covid and touche, differences are statistically non-significant (p > 0.1). (Sec.~\ref{appendix_tournament_llm})}
\label{table/appendix_tournament_llm}
\end{table}

\label{appendix_tournament_llm}
We have also analyzed the performance of RankGPT~\cite{rankgpt}, listwise reranking with GPT3.5, with tournament sort. The performance difference of tournament sort with respect to the sliding window approach for w=20, s=10 were not significant, while the number of required forward passes to rank top-10 passages were the same (9) for both variants.

\end{document}